\documentclass[pre,aps,onecolumn,superscriptaddress,nofootinbib]{revtex4-1}

\usepackage{amssymb}
\usepackage{amsmath}
\usepackage{graphicx}
\usepackage{array}
\usepackage{mathtools}
\usepackage{amssymb}
\usepackage{verbatim}
\usepackage{color}
\usepackage{bm}
\usepackage{amsthm}
\usepackage[normalem]{ulem}
\usepackage[titletoc,title]{appendix}

\usepackage[colorlinks=true, urlcolor=blue, anchorcolor=blue, citecolor=blue,filecolor=blue,linkcolor=blue,menucolor=blue
]{hyperref}

\usepackage{color}

\newcommand{\bea}{\begin{eqnarray}}
\newcommand{\eea}{\end{eqnarray}}
\newcommand{\be}{\begin{equation}}
\newcommand{\ee}{\end{equation}}
\newcommand{\nn}{\nonumber}
\newcommand{\bee}{\begin{equation*}}
\newcommand{\eee}{\end{equation*}}

\def\Xint#1{\mathchoice
   {\XXint\displaystyle\textstyle{#1}}%
   {\XXint\textstyle\scriptstyle{#1}}%
   {\XXint\scriptstyle\scriptscriptstyle{#1}}%
   {\XXint\scriptscriptstyle\scriptscriptstyle{#1}}%
   \!\int}
\def\XXint#1#2#3{{\setbox0=\hbox{$#1{#2#3}{\int}$}
     \vcenter{\hbox{$#2#3$}}\kern-.5\wd0}}

\def\dashint{\Xint-}

\def\Xint#1{\mathchoice
   {\XXint\displaystyle\textstyle{#1}}%
   {\XXint\textstyle\scriptstyle{#1}}%
   {\XXint\scriptstyle\scriptscriptstyle{#1}}%
   {\XXint\scriptscriptstyle\scriptscriptstyle{#1}}%
   \!\int}
\def\XXint#1#2#3{{\setbox0=\hbox{$#1{#2#3}{\int}$}
     \vcenter{\hbox{$#2#3$}}\kern-.5\wd0}}

\def\dashint{\Xint-}
\usepackage{dsfont}

\usepackage{fixmath}
\newcommand{\vect}[1]{\mathbold {#1}} 

\begin{document}

\title{Full Statistics of Nonstationary Heat Transfer in the Kipnis-Marchioro-Presutti Model}
\author{Eldad Bettelheim}
\email{eldad.bettelheim@mail.huji.ac.il}
\affiliation{Racah Institute of Physics, Hebrew University of
Jerusalem, Jerusalem 91904, Israel}
\author{Naftali R. Smith}
\email{naftalismith@gmail.com}
\affiliation{Department of Solar Energy and Environmental Physics, Blaustein Institutes for Desert Research, Ben-Gurion University of the Negev, Sede Boqer Campus, 8499000, Israel}
\author{Baruch Meerson}
\email{meerson@mail.huji.ac.il}
\affiliation{Racah Institute of Physics, Hebrew University of
Jerusalem, Jerusalem 91904, Israel}

\begin{abstract}
We investigate non-stationary heat transfer in the Kipnis-Marchioro-Presutti (KMP) lattice gas model at long times  in one dimension when starting from a localized heat distribution. At large scales this initial condition can be described as a delta-function, $u(x,t=0)=W \delta(x)$. We characterize the process by the heat transferred to the right of a specified point $x=X$ by time $T$,
$$
J=\int_X^\infty u(x,t=T)\,dx\,,
$$
and study the full probability distribution $\mathcal{P}(J,X,T)$. The particular case of $X=0$ has been recently solved [Bettelheim \textit{et al}. Phys. Rev. Lett. \textbf{128}, 130602 (2022)]. At fixed $J$, the distribution $\mathcal{P}$ as a function of $X$ and $T$ has the same long-time dynamical scaling properties as the position of a tracer in a single-file diffusion. Here we evaluate $\mathcal{P}(J,X,T)$ by
exploiting the recently uncovered complete  integrability of the equations of the macroscopic fluctuation theory (MFT) for the KMP model and using the Zakharov-Shabat inverse scattering method. We also discuss asymptotics of $\mathcal{P}(J,X,T)$ which we extract from the exact solution and also obtain by applying two different perturbation methods directly to the MFT equations.

\end{abstract}
\maketitle

{%
        \hypersetup{linkcolor=blue}
        \tableofcontents
}

\section{Introduction}
\label{intro}

One fundamental problem of statistical mechanics
deals with full statistics of currents of matter or
energy in classical many-body systems away from thermodynamic
equilibrium.  This challenging problem  has continued to attract
attention of workers in the past two decades. There has been a remarkable progress
in determining
the current statistics for nonequilibrium steady states in simple
models of interacting particles \cite{Derrida2007, BlytheEvans, AppertRolland,Lecomte}. Nonstationary current fluctuations, however, proved to be much more difficult to capture \cite{DG2009a,DG2009b,KrMe,MS2013,MS2014,VMS2014,ZarfatyM}.

A convenient and widely used family of models for
studying the full statistics of currents is stochastic lattice
gases \cite{Spohn,Liggett,KL,Krapivskybook}. The   Kipnis-Marchioro-Presutti (KMP) model \cite{KMP} is an important member of 
this family of models.
The KMP model is considered a  prototypical model of energy transfer by diffusion: not in the least because, for this model,  the
Fourier's law of heat diffusion  at a coarse-grained level was proven rigorously  \cite{KMP}. The model's definition is quite simple: There are
immobile agents occupying the whole lattice and carrying continuous amounts of a scalar quantity which can be viewed as energy. At each (exponentially distributed) random move  the total
energy of a randomly chosen pair of
nearest neighbors is redistributed among them randomly  according to uniform distribution, and the process continues.
Because of its simplicity the KMP model has become a paradigmatic model in the studies of fluctuations, including large deviations, of heat transfer in classical macroscopic nonequilibrium systems \cite{Bertini2005,BodineauDerrida,DG2009b,Lecomte,HurtadoGarrido,KrMe,
Pradosetal,MS2013,Peletier,JonaLasinioreview,ZarfatyM,Spielberg,Guttierez,Frassek,Grabsch,BSM1}.

\begin{figure}[ht]
\includegraphics[width=0.35\textwidth,clip=]{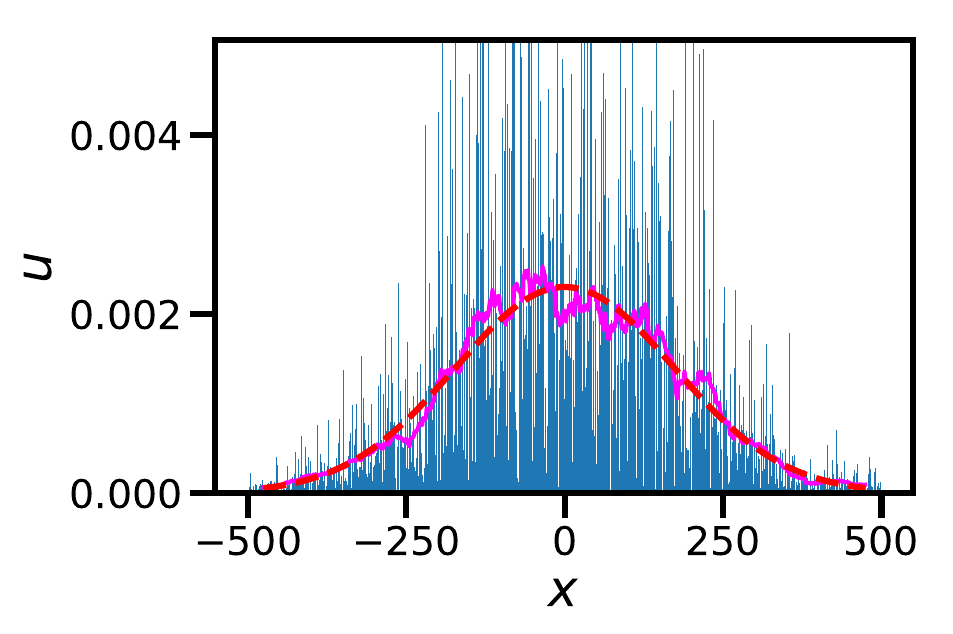}
\caption{A Monte-Carlo simulation of the microscopic KMP model. Initially only one agent, at $x=0$, had a nonzero energy $W=1$. Plotted is the simulated energy profile $u$ as a function of $x$ at time $t=3 \times {10}^4$ (bars), its spatial average over each 50 consecutive lattice sites (solid line), and the theoretical mean-field profile~(\ref{meanfield}) (dashed line).
}
\label{fig:sim}
\end{figure}

To set up our ideas, let us consider the KMP model on a one-dimensional lattice. Suppose that only one agent
has energy $W$ (or a few neighboring agents have total energy $W$)
at $t=0$. Due to the interactions with the neighbors, the energy will spread in a random fashion throughout the system. Looking for universality, we will focus on times much longer than the inverse rate of the elemental energy exchange between the two neighbors (equal to $1/2$),  and on distances much larger than the lattice constant (equal to $1$). As it was shown already in the original paper of KMP \cite{KMP}, the  mean, or expected, coarse-grained temperature $\bar{u}(x,t)$
obeys the heat diffusion equation
\begin{equation}\label{mfequation}
\partial_t \bar{u} (x,t) = \partial_x^2 \bar{u}(x,t)\,.
\end{equation}
The initial coarse-grained temperature profile is a delta-function,
\begin{equation}\label{mfinitial}
\bar{u} (x,t=0) = W \delta(x) ,
\end{equation}
so the mean-field solution for the coarse-grained temperature is
\begin{equation}\label{meanfield}
\bar{u}(x,t) = \frac{W}{\sqrt{4\pi t}}\,\exp\left(-\frac{x^2}{4t}\right)\,.
\end{equation}
However, in individual realizations of the stochastic process, the coarse-grained temperature $u(x,t)$ will of course fluctuate around the mean-field solution. These fluctuations are exemplified by Fig.~\ref{fig:sim} which shows a  snapshot, and a coarse-grained version of it, of a Monte Carlo simulation of the microscopic KMP model in this setting.  To characterize fluctuations of the heat transfer, we have recently evaluated \cite{BSM1} the full probability distribution $\mathcal{P}_0(J_0,T)$ of the total amount of heat $J_0$, observed at time $t=T\gg 1$ on the right half-line:
\begin{equation}\label{J0definition}
J_0=\int_0^{\infty} u(x,t=T) \,dx\,.
\end{equation}

Our work \cite{BSM1} continued a line of studies of large deviations of non-stationary mass or energy transfer in different diffusive lattice gas models \cite{DG2009a,DG2009b,KrMe,MS2013,MS2014,VMS2014}. In parallel, there has been much interest in the large deviations of the position of a tracer (or a tagged particle) in single-file systems such as the simple symmetric exclusion process (SSEP)
\cite{KMS1,KMS2,IMS1,IMS2,Grabsch,Poncet1,Poncet2,Illien}. (The SSEP is another well-known diffusive lattice gas model \cite{Spohn,KL,Schuetz,Chou}, where each particle can hop with equal probabilities to one of the adjacent lattice sites, but a hop to a site occupied by another particle is forbidden.) However, prior to our work \cite{BSM1},  the full statistics of the corresponding quantities were determined  only for \emph{annealed} initial conditions, which assume a ``pre-thermalization" of the lattice gas (or of the regions of $x<0$ and $x>0$ thereof at different densities or temperatures)
\cite{DG2009a,DG2009b,IMS1,IMS2,Grabsch}, see also the more recent Ref. \cite{MMS22}.

The evaluation of the full distribution  $\mathcal{P}_0(J_0,T)$ for the quenched (that is deterministic) initial condition (\ref{mfinitial}) was achieved in Ref. \cite{BSM1} by employing the macroscopic fluctuation theory (MFT)  \cite{JonaLasinioreview}, which will be introduced shortly,
and by uncovering and exploiting complete  integrability of the MFT equations  for the KMP model by the Zakharov-Shabat inverse scattering method \cite{Ablowitz,Novikov}. In this work we extend the approach of Ref. \cite{BSM1} to a more general characterization of the stochastic heat transfer in the KMP model. We specify an arbitrary point $x=X$ and evaluate the full time-dependent probability distribution $\mathcal{P}(J,X,T)$ of the total amount of heat,
\begin{equation}\label{Jdefinition}
J=\int_X^{\infty} u(x,t=T) \,dx\,
\end{equation}
observed at time $t=T\gg 1$ to the right of $x=X$. Because of the energy conservation in the KMP model, the distribution $\mathcal{P}(J,X,T)$  must have compact support:  $0<J<W$.
The mean, or expected value of $J$ at fixed $X$ readily follows from the deterministic solution~(\ref{meanfield}):
\begin{equation}\label{onehalf}
\bar{J} = \bar{J}(X,T) = \int_{X}^{\infty} \bar{u}(x,T)\,dx = \frac{W}{2} \text{erfc}\left(\frac{X}{\sqrt{4 T}}\right)\,,
\end{equation}
where $\text{erfc}\, z =1-\text{erf} \,z$, and $\text{erf} \,z$ is the error function. For $X=0$ one obtains $\bar{J}=W/2$. The main objective of this work is to determine the full statistics of fluctuations of $J$ around the mean value $\bar{J}$.

The presence of a new parameter $0<X<\infty$ allows for a more detailed characterization of the system, as we show here.  Importantly, when the position $x=X$ is varied at fixed $J$, it becomes an analog of the position of a tracer in single-file systems such as the SSEP \cite{Grabsch, IMS1, IMS2}. Indeed, in a single-file system, the total number of particles to the right of the tracer position $x=X$ does not change with time, and this can be mimicked by keeping $J=\text{const}$ in the KMP model. As a result, as we discuss below, the probability distribution $\mathcal{P}(J,X,T)$  as a function of $X$  at $J=\text{const}$ has the same dynamical scaling properties as the probability distribution of the position $x=X$ of a tracer in single-file systems, such as the SSEP.

Like many previous studies of the statistics of energy transfer \cite{DG2009b,KrMe,MS2013,MS2014,VMS2014,BSM1,MMS22} and of the tracer position \cite{KMS1,KMS2}, we will  employ the MFT \cite{JonaLasinioreview}:  a weak-noise theory based on a saddle-point evaluation of the exact path integral for the fluctuational hydrodynamics of diffusive lattice gases. The MFT is a variant of the optimal fluctuation method (OFM) which goes back to Refs. \cite{Halperin,Langer,Lifshitz}. The OFM has been also applied to turbulence \cite{turb1}, stochastic reactions \cite{EK,MS2011}, non-equilibrium surface growth and related models \cite{Fogedby,KK,MKV,HMS} and to many other systems. A full list of references on different applications of the OFM would exceed a hundred.

Fluctuational hydrodynamics of diffusive lattice gases \cite{Spohn,KL} is a coarse-grained description of the lattice gas on length scales much larger than the lattice constant 1, and time scales much larger than the inverse rate 1/2 of the energy exchange between the neighboring agents. The  corresponding small parameter in our problem is $1/\sqrt{T}\ll 1$ \cite{BSM1}.  For a general one-component diffusive lattice gas, the fluctuational hydrodynamics has the form of the stochastic partial differential equation \cite{Spohn,KL}
\begin{equation}
    \label{Langevingen}
    \partial_t u = \partial_x \left[D(u) \partial_x u+\sqrt{\sigma(u)}\,\eta\right]\,,
\end{equation}
where $u(x,t)$ is the coarse-grained density or temperature, $D(u)$ is the diffusion (or heat diffusion) coefficient, $\sigma(u)$ is the gas compressibility, and  $\eta(x,t)$ is a delta-correlated Gaussian noise,
$ \left<\eta(x,t)\right> = 0$ and
$\quad \left<\eta(x,t)\eta(x',t')\right> = \delta(x-x')\delta(t-t')$. For the KMP model, where $u(x,t)$ is the coarse-grained temperature, one obtains
$D=1$ and $\sigma(u) = 2u^2$, and Eq.~(\ref{Langevingen}) becomes
\begin{equation}
    \label{Langevin}
    \partial_t u = \partial_x \left(\partial_x u+\sqrt{2} u \eta\right)\,.
\end{equation}
Without the noise term Eq.~(\ref{Langevin}) becomes the heat diffusion equation~(\ref{mfequation}).

As mentioned above, the MFT relies on a saddle-point evaluation of the path integral for the stochastic process, described by Eq.~(\ref{Langevingen}). In our problem, the small parameter of the saddle-point evaluation is again $1/\sqrt{T}$. The saddle-point evaluation leads to a constrained optimization problem, where the (integral) constraint comes from conditioning the process on the specified transferred heat (\ref{Jdefinition}) subject to the initial condition (\ref{mfinitial}). The integral constraint (\ref{Jdefinition}) is conveniently taken into account via a Lagrange multiplier \cite{DG2009b}. We will present a complete Hamiltonian formulation of the MFT problem shortly. The solution of this problem describes the \emph{optimal path} of the process: the most likely time histories of the coarse-grained temperature  field and of the noise field which dominate the probability distribution $\mathcal{P}(J,X,T)$ that we are after.

The MFT, being stripped of unnecessary microscopic details, is an appealing tool for studying large deviations in lattice gases on large scales and at long times. However, the MFT equations -- two coupled  nonlinear partial differential equations -- are usually very hard to solve exactly, without reliance on additional small or large parameters. Ref. \cite{BSM1} reported a first major advance in this area of statistical mechanics by finding an exact solution to the MFT problem for the distribution  $\mathcal{P}(J,X,T)$ in the particular case of $X=0$. As already mentioned above, the exact solution came from uncovering and exploiting complete  integrability of the MFT equations   by the Zakharov-Shabat inverse scattering method. A second major advance in this area has been recently reported in Ref. \cite{MMS22}. The authors employed the MFT for determining the full statistics of
$J$ defined in Eq.~(\ref{Jdefinition}) for the SSEP with a step-like annealed initial condition. Earlier this problem was exactly solved (by a Bethe ansatz) in its microscopic formulation,  and the long-time asymptotic was extracted,  in Ref. \cite{DG2009b}. The authors of Ref. \cite{MMS22} reproduced this asymptotic by combining a generalization of the canonical Cole-Hopf transformation with the inverse scattering method.

The recent works \cite{BSM1} and \cite{MMS22} (and Refs. \cite{JKM,KLD1,KLD2} where complete integrability was uncovered \cite{JKM} and exploited \cite{KLD1,KLD2} for the determination of full short-time statistics
of the interface height as described by the Kardar-Parisi-Zhang (KPZ) equation \cite{KPZ}), have shown a great potential for the combination of optimal fluctuation methods, such as the MFT, with the inverse scattering method for obtaining exact solutions for different large-deviation statistics.
The present work is a natural next step in exploring this potential.

In Sec. \ref{MFTdef} we present the formulation of the MFT problem for $\mathcal{P}(J,X,T)$ \cite{DG2009b,MS2013} and briefly discuss the scaling behavior of $\mathcal{P}(J,X,T)$ and its relation with the universal scaling behavior of the tracer statistics in single-file processes.  In Sec.  \ref{solutionMFT} we solve the MFT problem by the inverse scattering method. In Sec. \ref{asymptot} we discuss some asymptotic limits of $\mathcal{P}(J,X,T)$. Our results are briefly summarized in Sec. \ref{discussion}.
Some of the technical details of the calculations are given in the Appendices.

\section{Formulation of the MFT problem}
\label{MFTdef}

We rescale the variables $t$, $x$ and $u$ by $T$, $\sqrt{T}$ and $W/\sqrt{T}$, respectively. As a result, $X$ and $J$ become rescaled by $\sqrt{T}$ and by $W$, respectively.  The optimal path of the process is described by two coupled Hamilton's equations: partial differential equations for the rescaled temperature field $u(x,t)$ and the conjugate ``momentum density" field $p(x,t)$ (the latter describes the optimal history of the noise $\eta(x,t)$ conditioned on the transferred heat $J$ at given $X$).

We also introduce the (minus) momentum density  gradient field $v(x,t)=-\partial_x p(x,t)$. As was shown earlier \cite{DG2009b,MS2013,BSM1}, the MFT equations can be written as
\begin{eqnarray}
  \partial_t u &=& \partial_x (\partial_x u+2 u^2 v)\,, \label{d1} \\[1mm]
  \partial_t v &=& \partial_x (-\partial_{x} v+2 u v^2)\,. \label{d2}
\end{eqnarray}
To make this paper more self-contained, a derivation of Eqs.~\eqref{d1} and \eqref{d2} is given in Appendix \ref{app:MFT}.
The initial condition, after the rescaling, is
\begin{equation}\label{delta0}
u(x,t=0) = \delta(x)\,.
\end{equation}
The rescaled transferred heat at $t=T$ is constrained by the equation
\begin{equation}\label{current0}
\int_X^{\infty} u(x,t=1) \,dx  =J\,.
\end{equation}
The particular case $X=0$ was  solved in Ref. \cite{BSM1}. Being interested in $X\neq 0$ we can, without loss of generality, set $X>0$.
The constrained optimization problem yields, aside from Eqs.~(\ref{d1}) and~(\ref{d2}),
a second boundary condition in time \cite{DG2009b},
\begin{equation}
\label{vdelta}
v(x,t=1) = -\lambda \,\delta(x-X)\,,
\end{equation}
where $\lambda$ is a Lagrange multiplier, to be ultimately determined from Eq.~(\ref{current0}).
This MFT formulation possesses a nontrivial symmetry
\begin{equation}
\label{uvsymmetry}
v(x,t) = -\lambda \, u \left(X-x,1-t\right)\,,
\end{equation}
which generalizes the symmetry found for the particular case $X=0$ in Ref. \cite{BSM1}. The symmetry \eqref{uvsymmetry} is special to the delta-function initial condition \eqref{delta0}.

Once $u(x,t)$ and $v(x,t)$ are determined, the rescaled mechanical action can be evaluated \cite{DG2009b,KrMe,MS2013}:
\begin{eqnarray}
  s  = \int_0^1 dt \int_{-\infty}^\infty dx \,u^2 v^2. \label{action0}
\end{eqnarray}
This action determines the probability density ${\mathcal P}(J,X,T)$ up to a pre-exponential factor. In the original, dependent variables we have
\begin{equation}\label{scalingc}
\ln {\mathcal P}(J,X,T)\simeq
-\sqrt{T} \,s\left(\frac{J}{W},\frac{X}{\sqrt{T}}\right)\,,
\end{equation}
which describes the scaling behavior of ${\mathcal P}(J,X,T)$. As we will see in the following, at fixed $J/W$ and small $X/\sqrt{T}$, 
the rate function $s$ is quadratic in $X/\sqrt{T}$, describing Gaussian fluctuations, where the standard deviation of $X$  scales with time as $T^{1/4}$, that is sub-diffusively. This scaling coincides with the one universally observed for the tracer position in single file diffusion processes, see Ref. \cite{KMS1,KMS2} and references therein. This coincidence is of course not accidental. Indeed, the mathematical formulation of the MFT problem that we have just presented, is very similar to the MFT formulation
for the statistics of the position of a tracer \cite{KMS1,KMS2}. The differences are in the particular $u$-dependence of the multiplicative noise term in the Langevin equation (\ref{Langevin}), and in the delta-function initial condition (\ref{delta0}) (rather than a step-like one) that we adopted here. Neither of them alters the subdiffusive scaling $T^{1/4}$ of the standard deviation of $X$. Finite or large $X/\sqrt{T}$ corresponds to non-Gaussian large deviations which are already non-universal, and which we will explore.
At fixed $X$ and small $J/W$, the rate function $s$ is also quadratic and therefore describes Gaussian fluctuations around the mean, which are universal for a whole family of lattice gases. For a finite $J/W$ (large deviations) the distribution is non-Gaussian.

It was observed \cite{BSM1} that Eqs.~(\ref{d1}) and (\ref{d2})
coincide with the derivative nonlinear Shr\"{o}dinger equation (DNLSE) in imaginary time and space \cite{DNLSE}. With real time and space,  the DNLSE describes nonlinear electromagnetic wave propagation in plasmas and other systems \cite{KN}. The initial-value problem for the DNLSE is completely integrable by using the Zakharov-Shabat inverse scattering
method (ISM) adapted by Kaup and Newell for the DNLSE \cite{KN}.
The heat transfer statistics presents an additional difficulty, as it involves a boundary-value problem in time. Ref. \cite{BSM1} overcame this difficulty in the particular case $X=0$ by (i) making use of a shortcut (see, \textit{e.g.} Ref. \cite{Vivoetal}) which makes it possible to determine the rate function $s(J,X)$ even in the absence of the complete solutions $u(x,t)$ and $v(x,t)$, and (ii) exploiting the symmetry relation (\ref{uvsymmetry}). Here we extend the approach of Ref. \cite{BSM1} to an arbitrary $X$.

\section{Solution of the MFT problem}
\label{solutionMFT}

Let us outline the key ideas behind the approach that we shall use to solve the MFT problem: the ISM \cite{Ablowitz,Novikov}. A key property related to the integrability of Eqs.~\eqref{d1} and \eqref{d2} is the existence of a corresponding Lax pair. This means than Eqs.~\eqref{d1} and \eqref{d2} are equivalent to the compatibility condition of a system of two linear differential equations defined by a pair of matrices. For the latter system, one can define scattering amplitudes which depend on $u$ and $v$.
As explained below, one can then solve for the time evolution of the scattering amplitudes, which turns out to be very simple. By relating the fields $u$ and $v$ to the scattering amplitudes, at $t=0$ and $t=1$, the ISM will enable us to find the transferred heat $J=J(\lambda,X)$ which suffices for the calculation of $s=s(J,X)$.

\subsection{Scattering amplitudes and their dynamics}

Adapting the derivation of Kaup and Newell \cite{KN} to imaginary time and space, we consider the linear system
\be
\label{psiODEs}
\begin{cases}
\partial_{x}\vect{\psi}(x,t,k)=U(x,t,k)\vect{\psi}(x,t,k)\,,\\[1mm]
\partial_{t}\vect{\psi}(x,t,k)=V(x,t,k)\vect{\psi}(x,t,k)\,,
\end{cases}
\ee
where $\vect{\psi}(x,t,k)$ is a column vector of dimension 2, the matrices
\bea
U(x,t,k)&=&\begin{pmatrix}
- i  k/2 & - i   v\sqrt{  i  k} \\[1mm]
- i  u\sqrt{ i  k} &   i  k/2 \\
\end{pmatrix}, \\[1mm]
V(x,t,k)&=&\begin{pmatrix} k^2/2- i  kuv & - i (\sqrt{ i  k})^3v+ i  \sqrt{ i  k } \,\partial_x v- i  \sqrt{ i  k} 2 v^2u \\[1mm]
- i (\sqrt{ i  k})^3u+ i  \sqrt{ i  k }\, \partial_x u- i  \sqrt{ i  k} 2 u^2v &   -k^2/2+ i  kuv \\
\end{pmatrix}
\eea
constitute the Lax pair, and $k$ is a spectral parameter.
It is now straightforward to check that the compatibility condition $\partial_{t} \partial_{x}\vect{\psi} = \partial_{x}\partial_{t}\vect{\psi}$, which corresponds to
\begin{equation}
\label{eq:compatibilityUV}
\partial_{t}U-\partial_{x}V+\left[U,V\right] = 0,
\end{equation}
is indeed equivalent to Eqs.~\eqref{d1} and \eqref{d2}.

It is useful to define the matrix $\mathcal{T}(x,y,t,k)$ as the $x$-propagator of the system \eqref{psiODEs}, namely,  the solution to the equation
\be
\label{eq:dTdx}
\partial_x \mathcal{T}(x,y,t,k)=U(x,t,k)\mathcal{T}(x,y,t,k)
\ee
with the initial condition $\mathcal{T}(x,x,t,k )=I$ (the identity matrix).
At $x \to \pm \infty$, the fields $u(x,t)$ and $v(x,t)$ vanish, and  the matrix $U$ takes the simple form,
\be
U(x\to\pm\infty,t,k)=\begin{pmatrix}-ik/2 & 0\\
0 & ik/2
\end{pmatrix} \,.
\ee
As a result, the entries of $\mathcal{T}(x,y,t,k)$ oscillate as a function of $x$ and $y$ with wave numbers 
whose magnitude is $k/2$ at $x\to\pm\infty$, $y\to\pm\infty$.
Therefore, it is natural to define the full-space propagator $G(t,k)$ as follows:
\be
\label{eq:Ttkdef}
G(t,k)=\lim_{\begin{array}{c}
x\to\infty\\
y\to-\infty
\end{array}}\begin{pmatrix}e^{ i  kx/2} & 0\\
0 & e^{- i  kx/2}
\end{pmatrix}
 \mathcal{T}(x,y,t,k)\begin{pmatrix}e^{- i  ky/2} & 0\\
0 & e^{ i  ky/2}
\end{pmatrix}.
\ee
The matrix $G(t,k)$ plays an important role in the ISM, because its ($u$- and $v$-dependent) entries are the scattering amplitudes of the system \eqref{psiODEs}.
The time evolution of $G(t,k)$ is remarkably simple. Indeed, one finds that the time evolution of the matrix $\mathcal{T}(x,y,t,k)$ satisfies:
\be
  \partial_t \mathcal{T}(x,y,t,k) = V (x,t,k)\mathcal{T}(x,y,t,k)
   - \mathcal{T}(x,y,t,k)V(y,t,k) \, .
   \label{eq:dTdt}
\ee
As one can verify directly, Eq.~\eqref{eq:dTdt} is compatible with \eqref{eq:dTdx} (\textit{i.e.} $\partial_{t} \partial_{x}\mathcal{T} = \partial_{x}\partial_{t}\mathcal{T}$) due to Eq.~\eqref{eq:compatibilityUV}.
The matrix $V(x,t,k)$ too becomes very simple in the limit $x\to\pm\infty$,
\be
\label{eq:Vlim}
V(x\to\pm\infty,t,k)=\frac{k^{2}}{2}\begin{pmatrix}1 & 0\\
0 & -1
\end{pmatrix} \, .
\ee
Plugging Eq.~\eqref{eq:Vlim} into \eqref{eq:dTdt}, one finds the time evolution of $\mathcal{T}(x\to \infty, y \to -\infty, t,k)$ which in turn, using \eqref{eq:Ttkdef}, yields that of $G(t,k):$
\begin{align}
\label{eq:TtimeDependence}
G(t,k)=\begin{pmatrix}a(t,k) & \tilde b(t,k) \\[1mm]
b(t,k) & \tilde a(t,k) \\
\end{pmatrix}=\begin{pmatrix}a(0,k) & \tilde b(0,k)e^{k^2t} \\[1mm]
b(0,k)e^{-k^2t} & \tilde a(0,k) \\
\end{pmatrix}
\end{align}
In Eq.~\eqref{eq:TtimeDependence} we have introduced a notation for the matrix elements of $G(t,k)$.

\subsection{Solving the scattering problem at $t=0$ and $t=1$}

Let us find the matrix $\mathcal{T}(x,y,0,k) $ at $t=0$. This is achieved by solving Eq.~\eqref{eq:dTdx} at $t=0$, which reads (using  $u(x,0)=\delta(x)$)
\be
\label{Teqt0}
\partial_{x}\mathcal{T}\left(x,y,0,k\right)=\begin{pmatrix}-\frac{ik}{2} & -iv\left(x,0\right)\sqrt{ik}\\[1mm]
-i\delta\left(x\right)\sqrt{ik} & \frac{ik}{2}
\end{pmatrix}\mathcal{T}\left(x,y,0,k\right)\,.
\ee
The general solution to Eq.~\eqref{Teqt0} is found by solving it in the regions $x<0$ and $x>0$ and then matching the solutions at $x=0$. Then, taking into account the initial condition $\mathcal{T}(x,x,t,k )=I$, we obtain  (see Appendix \ref{app:t0} for details)
\be
\label{eq:Txyt0}
\mathcal{T}(x,y,0,k)=\begin{cases}
\begin{pmatrix}e^{-ik(x-y)/2} & -i\sqrt{ik}\,e^{-ik(x-y)/2}I_{v}(x,y)\\[1mm]
0 & e^{ik(x-y)/2}
\end{pmatrix}\,, & xy>0\,,\\
\\
\begin{pmatrix}e^{ik(y-x)/2}\left[1\pm ikI_{v}(x,0)\right] & -i\sqrt{ik}e^{ik(y-x)/2}\left[I_{v}(x,y)\pm ikI_{v}(0,y)I_{v}(x,0)\right]\\[1mm]
\pm i\sqrt{ik}e^{ik(x+y)/2} & e^{ik(x-y)/2}\pm ike^{ik(x+y)/2}I_{v}(0,y)
\end{pmatrix}\,, & xy<0\,,
\end{cases}
\ee
 where
\be
 I_v(x,y)=\int_y^x v(z,0) e^{ i  k(z-y)}dz,
\ee
and in the second case in \eqref{eq:Txyt0}, the sign $\pm $ is the same as the sign of $y.$
We now compute $G(0,k)$, by plugging Eq.~\eqref{eq:Txyt0} into \eqref{eq:Ttkdef}, with the result:
\be
\label{T(t=0)}
G(0,k)=\begin{pmatrix}1- i  kQ_{+}(k) & - i \sqrt{ i  k}\left[Q(k)- i  kQ_{-}(k)Q_{+}(k)\right]\\[1mm]
- i \sqrt{ i  k} & 1- i  kQ_{-}(k)
\end{pmatrix}
\ee
in terms of $Q_{\pm}(k)$ which
are the Fourier transforms of $v(z,0)$ restricted to $z>0$ and $z<0$, respectively, and of their sum:
\be
\label{Qdef}
Q_-(k)\!=\!\int_{-\infty}^0 \! v(z,0) e^{ i  kz}dz, \quad Q_+(k)=\int_{0}^\infty \! v(z,0) e^{ i  kz}dz\,, \quad Q(k) = Q_+(k) + Q_-(k) \, .
\ee

It is useful to compare this result to the one obtained at $t=1$.
Here we have $v(x,1)=  - \lambda\delta(x-X)$, which enables us to solve Eq.~\eqref{eq:dTdx} at $t=1$.
The solution is very similar to that at $t=0$, and one gets
\bea
\label{Teqt1}
&&\mathcal{T}(x,y,1,k) \nn\\[2mm]
&&\qquad=
\begin{cases}
\begin{pmatrix}e^{-ik(x-y)/2} & 0\\[1mm]
-i\sqrt{ik}\,e^{ik(x-y)/2}I_{u}(x,y) & e^{ik(x-y)/2}
\end{pmatrix}\,, & \left(x-X\right)\left(y-X\right)>0\,,\\
\\
\begin{pmatrix}e^{-ik(x-y)/2}\pm\lambda ike^{ik(x+y-2X)/2}I_{u}(X,y) & \pm i\lambda\sqrt{ik}e^{-ik(x+y-2X)/2}\\[1mm]
-i\sqrt{ik}e^{ik(x-y)/2}\left[I_{u}(x,y)\pm\lambda ikI_{u}(X,y)I_{u}(x,X)\right] & e^{ik(x-y)/2}\left[1\pm\lambda ikI_{u}(x,X)\right]
\end{pmatrix}\,, & \left(x-X\right)\left(y-X\right)<0\,,
\end{cases}
\eea
where
\be
I_{u}(x,y)=\int_{y}^{x}u(z,1)e^{- i  k(z-y)}dz,
\ee
and in the second case in \eqref{Teqt1}, the sign $\pm $ is to be taken as the opposite of the sign of $y-X.$
Similarly to the $t=0$ case, we now compute
\be
 \label{T(t=1)}
G(1,k)=\begin{pmatrix}1+\lambda ikR_{+}(k) & +i\lambda\sqrt{ik}\,e^{ikX}\\[1mm]
-i\sqrt{ik}\,e^{-ikX}\left[R(k)+\lambda ikR_{-}(k)R_{+}(k)\right] & 1+\lambda ikR_{-}(k)
\end{pmatrix}
\ee
where
\be
R_{+}(k)=\int_{-\infty}^{X}u(z,1)e^{-ikz}dz,\quad R_{-}(k)=\int_{X}^{\infty}u(z,1)e^{-ikz}dz,\quad R(k)=R_{+}(k)+R_{-}(k)\,.
\ee

\subsection{Recovering the rate function $s(J,X)$ from the scattering amplitudes}

In this section we calculate the rate function $s(J,X)$ by relating the scattering amplitudes at $t=0$ and $t=1$. We find that several cases must be considered.

(i) In the simplest case, in which $\lambda$ is sufficiently small, the derivation closely follows that of \cite{BSM1}. This case completely covers the case $X=0$.
(ii) For $X \ne 0$, and larger values of $\lambda$, technical problems are encountered when evaluating some of the integrals that appear in the solution in case (i). These are fairly simple to circumvent, by deforming the integration contours in the complex plane. This completely covers the case $X < \sqrt{8}$.
(iii) For $X > \sqrt{8}$ a third regime, of very large $\lambda$, exists. This regime is rather interesting as it involves multiple solutions at a given $\lambda$. That is, there is a regime of $\lambda$'s for which $s$ and $J$ are multi-valued functions of $\lambda$. Nevertheless, $s$ remains a single-valued, analytic function of $J$, unlike in many other systems in which additional branches of the rate function lead to singularities which have the character of dynamical phase transitions
\cite{ZarfatyM, Baek15, JKM, Baek17, Baek18,  Shpielberg2016, Spielberg, LeDoussal2017, SKM2018}.
As we explain below, this rather unusual behavior comes about concurrently with a non-convexity of $s(J,X)$ as a function of $J$ for $X > \sqrt{8}$.

\subsubsection{Full solution for $X=0$, and for $X>0$ with sufficiently small $\lambda$}

Comparing the upper-right elements of $G(0,k)$ from Eq.~(\ref{T(t=0)}) and of $G(1,k)$ from Eq.~(\ref{T(t=1)}), using Eq.~\eqref{eq:TtimeDependence}, we obtain
\be
\label{eq:Qpm}
ikQ(k)-ikQ_{-}(k)\times ikQ_{+}(k)=-\lambda ik\,e^{ikX-k^{2}} \, .
\ee
Eq.~\eqref{eq:Qpm} provides a key step towards the solution to our problem because, after solving it, $J=J(\lambda,X)$ is obtained from $Q_+(0)$ with the use of the symmetry \eqref{uvsymmetry}, and this, as shown below, suffices 
to determine of the rate function $s(J,X)$.
We now turn to the solution to Eq.~\eqref{eq:Qpm}.
We begin by completing the squares in Eq.~\eqref{eq:Qpm}, by writing
\begin{align}
\label{basically the result}
 i  kQ_\pm(k)=1-e^{\Phi_\pm(k)-\Phi_\pm(0)  }\,,
\end{align}
where the constant $-\Phi_\pm(0)$ in the exponent is found by requiring that $Q_\pm(k)$ is regular at the origin.
This equation turns Eq.~\eqref{eq:Qpm} into
\begin{align}
\label{eq:vpmMpm}
e^{\Phi_+(k)+\Phi_-(k)-\Phi_+(0)-\Phi_-(0)}=1 + i \lambda    k   e^{ikX-k^2},
\end{align}
which has the solution
\be
\label{Phi}
\Phi_{\pm}(k)= \pm\int_{-\infty}^{\infty}\frac{\ln\left(1 +  i \lambda k'e^{ik'X-k'^{2}}\right)}{k'-k\mp i 0^{+}}\frac{dk'}{2\pi i } \, .
\ee
Eq.~\eqref{Phi} is derived by  noting that $Q_\pm(k)$ are analytic in the upper and lower half complex $k$ plane, respectively, and are well-behaved when $k$ is allowed to reach infinity through the respective half-planes%
\footnote{These properties of $Q_\pm(k)$ follow from their definitions \eqref{Qdef} together with the expected Gaussian decay of $v(z \to \pm \infty,0)$ and boundedness of $v(z \to \pm 0,0)$.}.
We then use the well-known decomposition $f(k) = f_+(k) + f_-(k)$ of a general function $f(k)$ into functions analytic in the upper and lower half-planes, $f_\pm(k)$, respectively, given by
\be
\label{fpmDecomposition}
f_\pm(k)=\int \frac{f(k')}{k'-k\mp i 0^+}\frac{dk'}{2\pi i} \, .
\ee
The decomposition \eqref{fpmDecomposition} is applied to the logarithm of Eq. (\ref{eq:vpmMpm}),
leading to Eq.~\eqref{Phi}.

Taking the derivative of Eq. (\ref{basically the result}) with respect to $k$, setting $k$ to $0$ and dividing by $i$ we obtain:
\begin{align}
 \label{QpmNonPrincipal}
Q_{\pm}(0)=\pm\int_{-\infty}^{\infty}\frac{\ln\left(1+ i \lambda ke^{ i  kX-k^{2}}\right)}{(k \mp i 0^{+})^{2}}\frac{dk}{2\pi}
 \end{align}
which, using the Sokhotski--Plemelj formula, can be also written in terms of a principal value integral,
\begin{align}
\label{QpmPrincipalValue}
Q_{\pm}(0)=\pm\dashint_{-\infty}^{\infty} \frac{\ln\left(1+ i \lambda ke^{ i  kX-k^{2}}\right)}{k^{2}}\frac{dk}{2\pi}\mp\frac{\lambda}{2} \, .
\end{align}
Using Eqs.~\eqref{current0}, (\ref{uvsymmetry}) and \eqref{Qdef} alongside with the conservation law $\int_{-\infty}^{\infty} u(x,t)\,dx =1$, we determine $J=J(\lambda, X)$:
\bea
\label{jlambda}
J\left(\lambda,X\right)&=&\int_{X}^{\infty}u\left(x,t=1\right)\,dx=1-\int_{-\infty}^{X}u\left(x,t=1\right)\,dx
=1+\frac{1}{\lambda}\int_{0}^{\infty}v\left(x,0\right)\,dx \nn\\
&=&1+\frac{Q_{+}\left(0\right)}{\lambda}=1+\int_{-\infty}^{\infty}\frac{\ln\left(1+i\lambda ke^{ikX-k^{2}}\right)}{(k - i0^{+})^{2}}\frac{dk}{2\pi\lambda}
\eea
which can also be written as
\be
J\left(\lambda, X\right)=\dashint_{-\infty}^{\infty}\frac{\ln\left(1+ i \lambda ke^{ i  kX-k^{2}}\right)}{k^{2}}\frac{dk}{2\pi\lambda} - \frac{1}{2}\,.
\ee

In order to proceed, we utilize a useful shortcut, which enables us to obtain the rate function $s=s(J,X)$ given that we have calculated $J=J(\lambda,X)$. The shortcut makes use of the relation $\partial_{J}s=\lambda$, a property that follows from the fact that $J$ and $\lambda$ are conjugate variables, see \textit{e.g.} Ref. \cite{Vivoetal}.
This enables us to calculate $s(J,X)$ bypassing Eq.~(\ref{action0}) [which would require us to know the entire optimal path $u(x,t)$].
Using $J\left(\lambda,X\right)=1 + Q_{+}\left(0\right)/\lambda$ from Eq.~\eqref{jlambda}, we have
\be
\label{eq:dsdlambda}
\frac{\partial s}{\partial\lambda}=\frac{\partial s}{\partial J}\frac{\partial J}{\partial\lambda}=\lambda\frac{\partial J}{\partial\lambda}=\frac{\partial Q_{+}\left(0\right)}{\partial\lambda}-\frac{Q_{+}\left(0\right)}{\lambda}\,.
\ee
Using Eq.~\eqref{QpmPrincipalValue}, we integrate Eq.~\eqref{eq:dsdlambda} with respect to $\lambda$ to get
\begin{equation}
\label{slambdasimpler}
s(\lambda, X)=Q_{+}(0)+\int_{-\infty}^{\infty}\frac{\text{Li}_{2}\left(-i k \lambda e^{ikX-k^{2}}\right)}{2\pi k^{2}}\,dk\,+\frac{\lambda}{2}.
\end{equation}
where
$\text{Li}_{2}\left(z\right)=\sum_{k=1}^{\infty}z^{k}/k^2$ is the dilogarithm function, $Q_{+}(0)$ is given by Eq.~\eqref{QpmPrincipalValue}, and the integration constant was determined from  $s(\lambda=0,X)=0.$ Equations~\eqref{jlambda} and \eqref{slambdasimpler} give the rate function $s(J, X)$ in a parametric form.

\begin{figure}
\begin{center}
\includegraphics[width=0.6\textwidth]{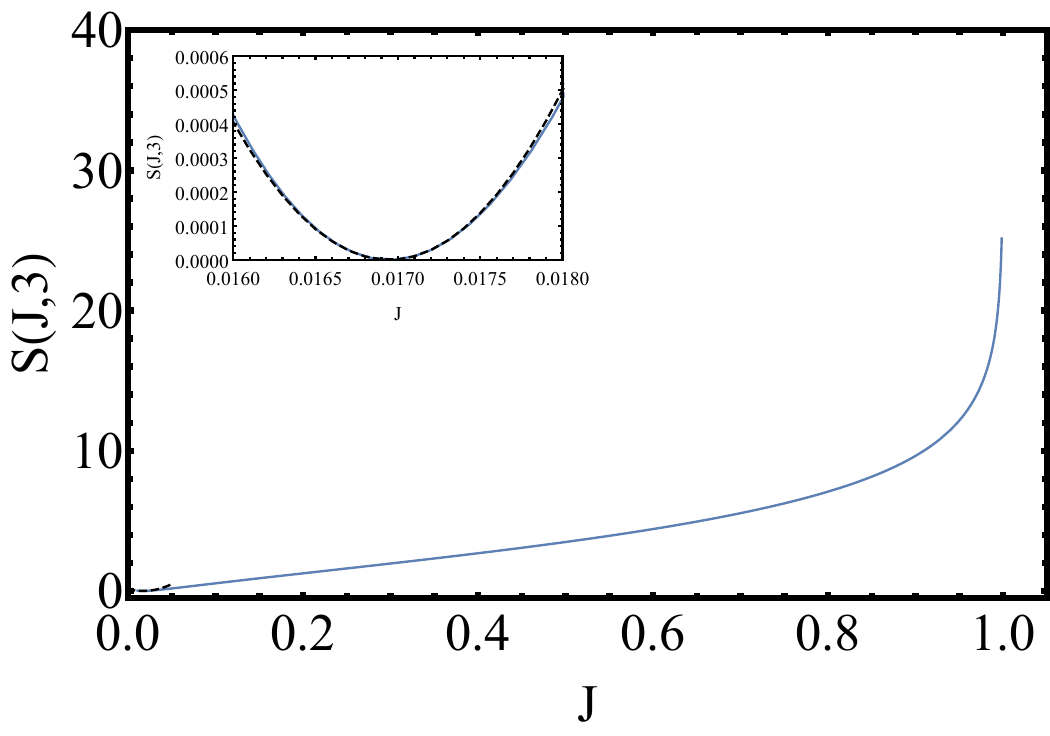}
\caption{Action $s(J,X=3)$ as a function of $J$ for $X=3$.
Dashed line is the asymptote at $J\to \bar{J} = \text{erfc}(3/2)/2 = 0.016947\dots$, given by Eqs.~\eqref{slinear} below.
The inset shows an enlargement of the asymptotic region.}
\label{svsj}
\end{center}
\end{figure}

\begin{figure}[ht]
\begin{center}
\includegraphics[width=0.4\textwidth,clip=]{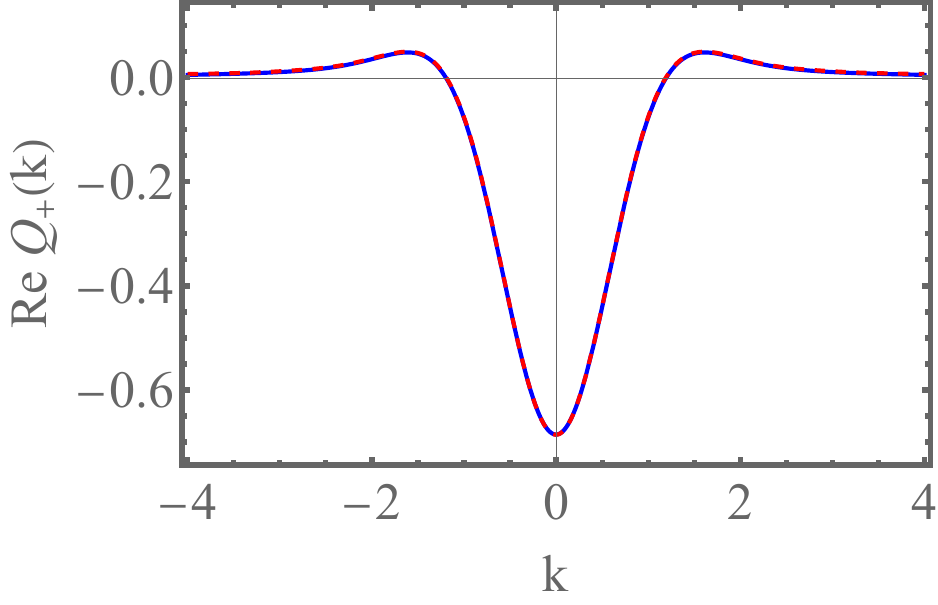}
\hspace{2mm}
\includegraphics[width=0.4\textwidth,clip=]{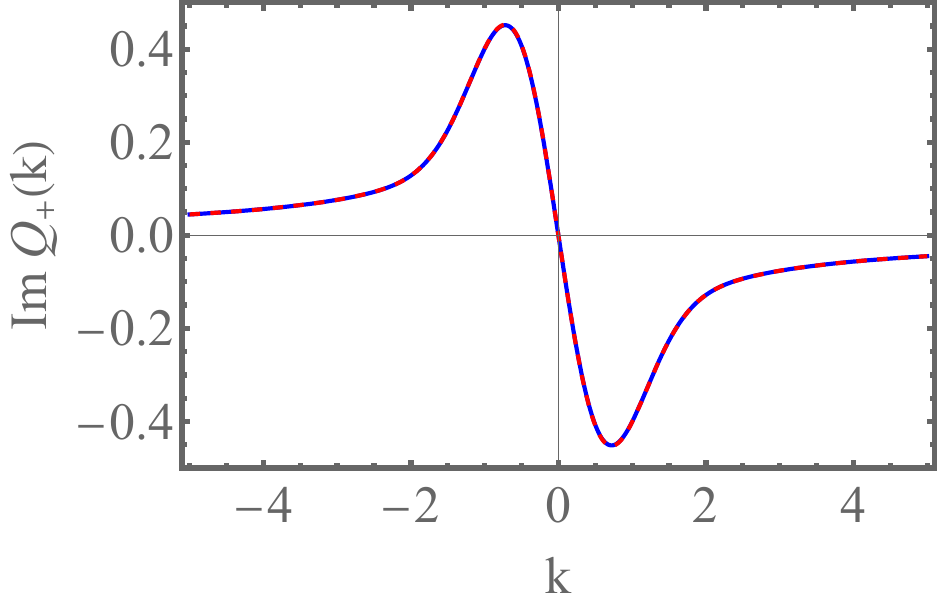}
\caption{Analytical results for the real and imaginary parts of $Q_+(k)$, described by Eqs. (\ref{basically the result}) and (\ref{Phi})  (solid lines), versus numerical results (dashed lines), for $\lambda=X=1$, corresonding to $J=0.31\dots$.}
\label{QplusX1}
\end{center}
\end{figure}

Figure \ref{svsj} shows $s(J,X)$ alongside with the asymptotic $J\to \bar{J}$, which corresponds to $\lambda \to 0$.  This asymptotic
can be obtained either from the exact rate function \eqref{jlambda} and \eqref{slambdasimpler}, or from a small-$\lambda$ perturbative expansion applied  directly to the MFT equations \cite{KrMe, BSM1}.
Figure  \ref{QplusX1} shows $\text{Re}\, Q_+(k)$ and $\text{Im}\,Q_+(k)$ versus $k$ at $X=1$.
This figure also shows the same quantities computed by solving  Eqs.~(\ref{d1}) and (\ref{d2}) numerically with a back-and-forth iteration algorithm \cite{CS}. The analytical and numerical curves are in good agreement.

We now deal with the question of the uniqueness of the solution. Let us denote $a_{+}\left(k\right)=a\left(k,0\right)$ and $a_{-}\left(k\right)=\tilde{a}\left(k,0\right)$ [using the notation introduced in \eqref{eq:TtimeDependence}]. From Eqs.~\eqref{T(t=0)} and \eqref{T(t=1)}, we have
\be
a_{\pm}\left(k\right)=1-ikQ_{\pm}\left(k\right)=1+\lambda ikR_{\pm}\left(k\right) \, .
\ee
Now we turn to Eq. \eqref{eq:Qpm} and to its solution, given by Eq.~\eqref{basically the result}. In terms of $a_{\pm}$ these two equations read
\bea
\label{apmEq}
&& 1-a_{+}\left(k\right)a_{-}\left(k\right) = -\lambda ik\,e^{ikX-k^{2}}, \\
&& a_{\pm}\left(k\right)=e^{\Phi_{\pm}(k)-\Phi_{\pm}(0)} , \label{adefPhi}
\eea
respectively.
However, there are more solutions to Eq.~\eqref{apmEq}.
Given any solution $a_{\pm}\left(k\right)$ to Eq.~\eqref{apmEq}, it is easily seen that, for any $k_1, k_2 \in \mathds{C}$,
where $k_1$ is in the upper half plane and $k_2$ in the lower half plane, \be
\label{aHat}
\hat{a}_{\pm}\left(k\right)=\left(\frac{k_2}{k_1}\frac{k-k_{1}}{k-k_{2}}\right)^{\pm1}a_{\pm}\left(k\right)
\ee
is also a solution, since only the product $a_{+}\left(k\right)a_{-}\left(k\right)$ enters in \eqref{apmEq}.
This transformation can be applied any number of times. Which of these solutions is the proper solution to the MFT equations, i.e., the solution which describes histories of $u$ and $v$ that are real, nonnegative and minimize the action $s$ constrained on a given $J$?  One useful test is based on the fact that the solutions for $u$ and $v$ must exhibit a Gaussian decay at $x \to \pm \infty$. This is because $u$ and $v$ are small there, and therefore the nonlinear terms in the MFT equations \eqref{d1} and \eqref{d2} are negligible, so these equations become the diffusion and anti-diffusion equations respectively. For localized initial conditions, the solutions are Gaussian at $x \to \pm \infty$ at all times. On the other hand, the factors which appear in the transformation \eqref{aHat} lead, in general, to an exponential rather than a Gaussian decay of $u(x,1)$ or $v(x,0)$ at $x \to \pm \infty$ due to the poles introduced by the Fourier transforms of these functions.
The exponential decay is avoided, if $k_{1}$ and $k_2$ solve the equation
 \begin{align}
 \label{kiEq}
1+i\lambda k_{i}e^{i k_{i}X- k_{i}^{2}}=0\ .
\end{align}
In this case the pole cancels out, and the inclusion of the extra factor in $a_\pm(k)$ does not alter the Gaussian decay at infinity. Another possibility is that $k_i$ is real (or infinitesimally below or above the real axis). These observations will be useful below.

\subsubsection{Solution for $X<\sqrt{8}$ at all $\lambda$}
\label{sec:xlesqrt2}

In the analysis up to now, we tacitly assumed that in the term $\ln\left(1+ i \lambda ke^{ i  kX-k^{2}}\right)$ branch cuts of the logarithm are never encountered in the integrals given above. This is indeed the case for $X=0$ and also at sufficiently small $\lambda$ for $X \ne 0$.  However, at larger $\lambda$,
 branch cuts are encountered at $X \ne 0$, i.e., for some $k \in \mathds{R}$,
\begin{align}
\label{R-condition}
1+  i  \lambda ke^{ i  kX-k^2} \in \mathds{R}_- \, .
\end{align}
The values of $k$ and $\lambda$ at which these branch cuts cross the real axis can be found by solving the equation $1+i\lambda ke^{ikX-k^{2}}=0$ for real $k$ and $\lambda.$
The solutions are $k = \pm k_{R,n}$, where
\begin{align}
\label{knsol}
&k_{R,n} =\frac{\left(n+\frac{1}{2}\right)\pi}{X}, \qquad n  = 0,1,2,\dots, \\
&\lambda_{c}\left(X,n\right) =\frac{\left(-1\right)^{n}X}{\left(n+\frac{1}{2}\right)\pi}\exp\left[\frac{\left(n+\frac{1}{2}\right)^{2}\pi^{2}}{X^{2}}\right] \, .
\end{align}
It follows that for each $n$ for which  $\lambda > \lambda_{c}(X,n)$, there is an additional branch cut crossing the real axis at $k= \pm k_{R,n}$.

\begin{figure}[ht]
\includegraphics[width=0.45\textwidth,clip=]{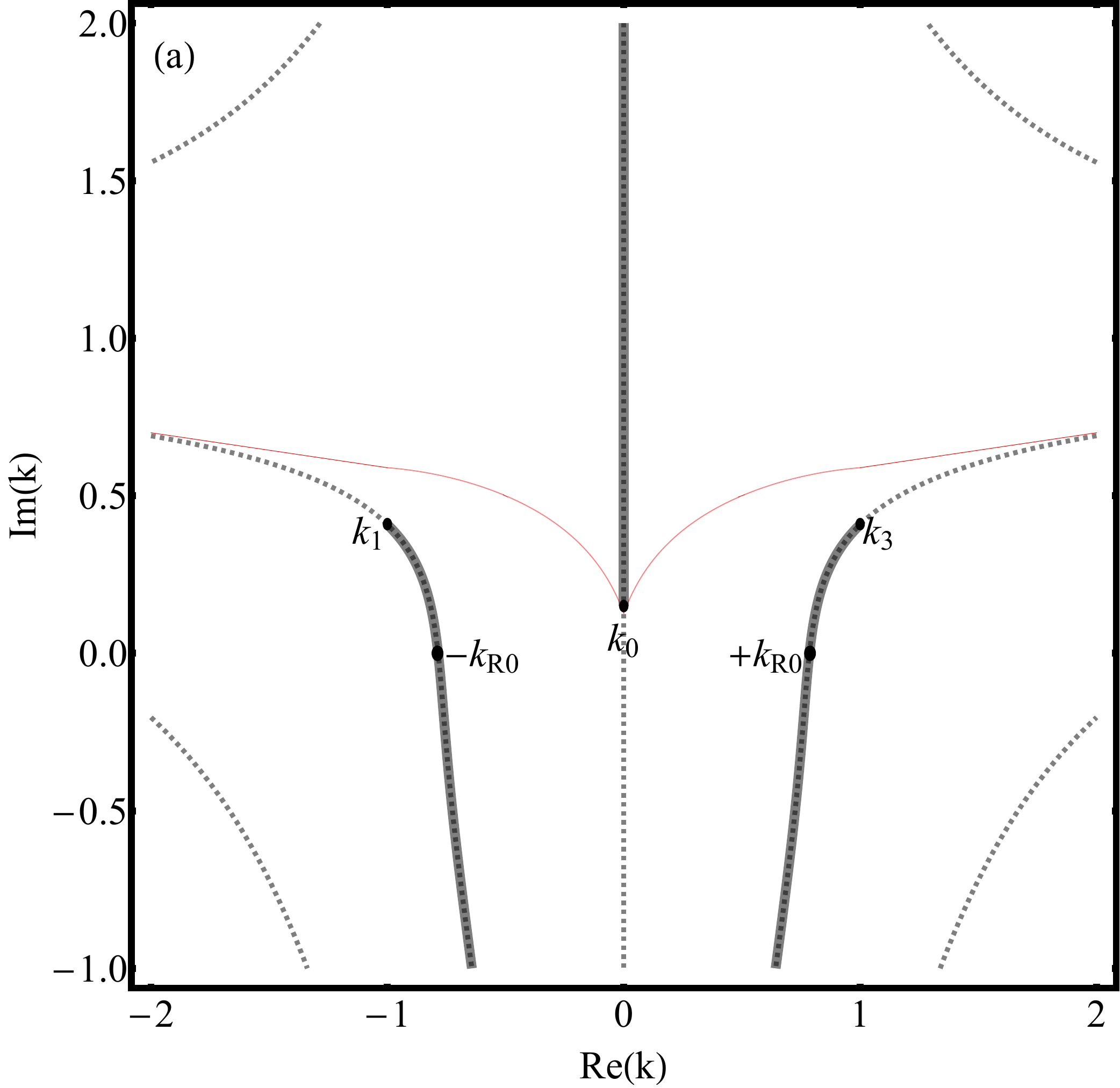}
\hspace{2mm}
\includegraphics[width=0.45\textwidth,clip=]{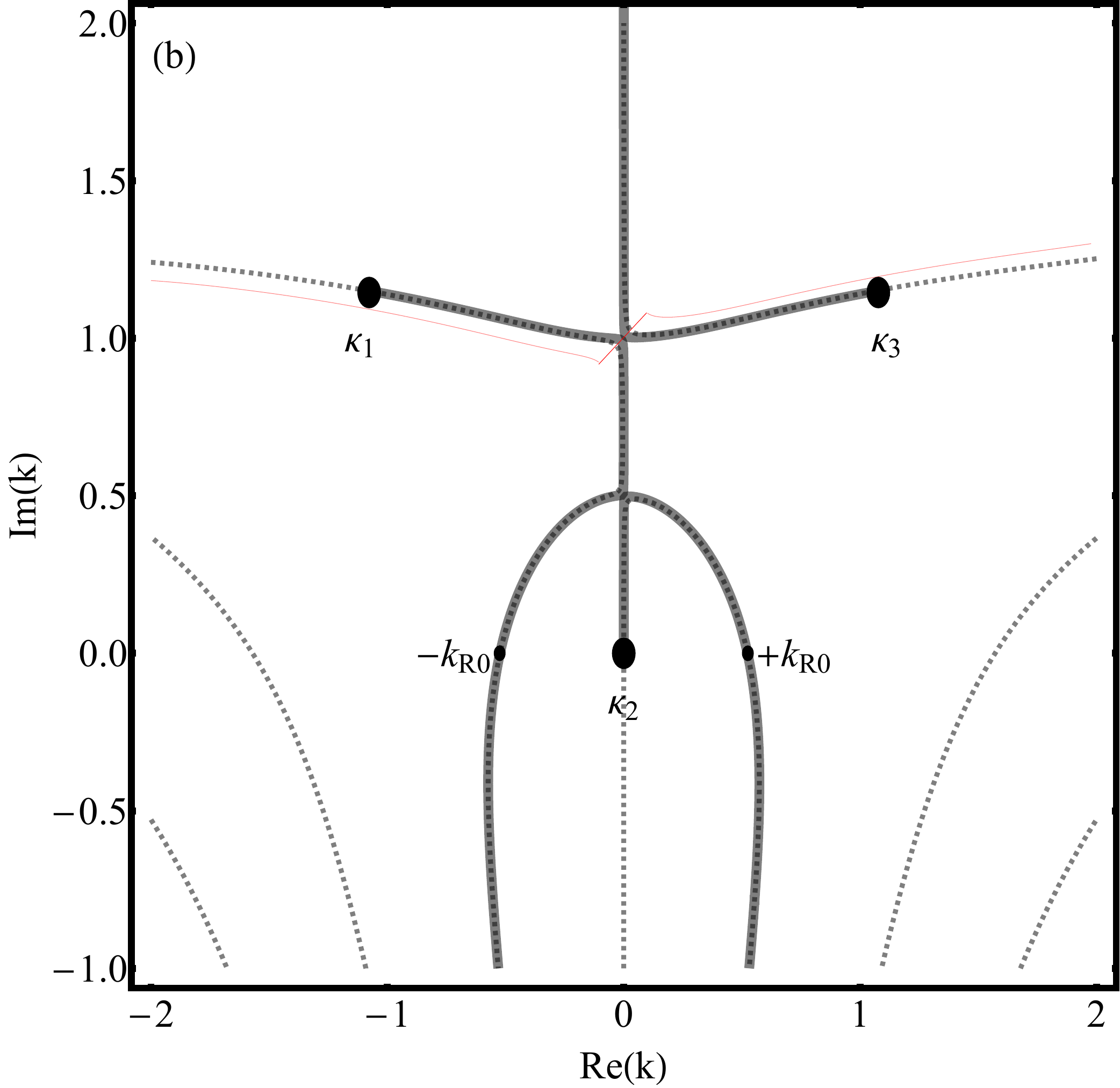}
\caption{Contour plot of $\arg i  ke^{ i  kX-k^{2}}$ for $\lambda=10$ and  $X=2$ (a) and $X=3$ (b), in the complex $k$ plane. Dotted and solid lines correspond to $ i  ke^{ i  kX-k^{2}}\in\mathds{R}_{\pm}$ respectively. The solid lines  are branch cuts of $\ln\left(1+ i \lambda ke^{ i  kX-k^{2}}\right).$  A convenient choice for integration contours are the red  lines while the points at which the branch cuts cross the real axis are denoted by $\pm k_{R0}$ in accordance with Eq. (\ref{knsol}). In panel (b) the branch points $\kappa_i$ are denoted, while in panel (a) three branch points are arbitrarily denoted as $k_i$, $i=0,1,3$. }
\label{figContourX}
\end{figure}

The branch cut crossing the real axis would cause $\Phi_\pm$ to be non-analytic as a function of $\lambda$ if the real axis would be retained as the contour of integration in Eq. (\ref{Phi}). This would lead to non-analyticity in all the physical quantities ($J$, $S$, etc.). To retain analyticity, one must change the contour of integration in Eq. (\ref{Phi}), and in
Eqs.~\eqref{QpmNonPrincipal}, \eqref{jlambda}, \eqref{slambdasimpler} stemming from it,
such as to avoid crossing any branch cut. Possible contours are shown in Fig. \ref{figContourX}, but any contour that does not cross the branch cuts would do.

\subsubsection{Solution for $X > \sqrt{8}$ at all $\lambda$}

At $X > \sqrt{8}$, it is impossible to avoid the branch cuts altogether at sufficiently large $\lambda$, see below. In fact, the analytical continuation of the formula of Eq. (\ref{Phi}) leads to a complex $J.$ We show these statements and obtain the proper solutions for any value $\lambda>0$ below (at $\lambda < 0$, the solution for the case $X < \sqrt{8}$ is valid here as well).

It is convenient at this point to use a different method to analytically continue the solutions from small $\lambda$ (as opposed to the one used for $X<\sqrt{8}$, which consisted of changing the contour of integration).   Let us first define, for the case $X<\sqrt{8}$, three critical values of $\lambda$, which will be important in the following:
\be
\lambda_{c0}=\lambda_{c}\left(X,0\right)<\lambda_{c1}=\frac{4e^{\frac{1}{2}+\frac{X^{2}
-X\sqrt{X^{2}-8}}{8}}}{X-\sqrt{X^{2}-8}}<\lambda_{c2}=\frac{4e^{\frac{1}{2}+\frac{X^{2}
+X\sqrt{X^{2}-8}}{8}}}{X+\sqrt{X^{2}-8}}\,.
\ee
At $\lambda=\lambda_{c1}$ and $\lambda=\lambda_{c2}$, the number of purely imaginary solutions to the equation \eqref{kiEq} changes.

\begin{figure}[ht]
\begin{center}
\includegraphics[width=0.4\textwidth,clip=]{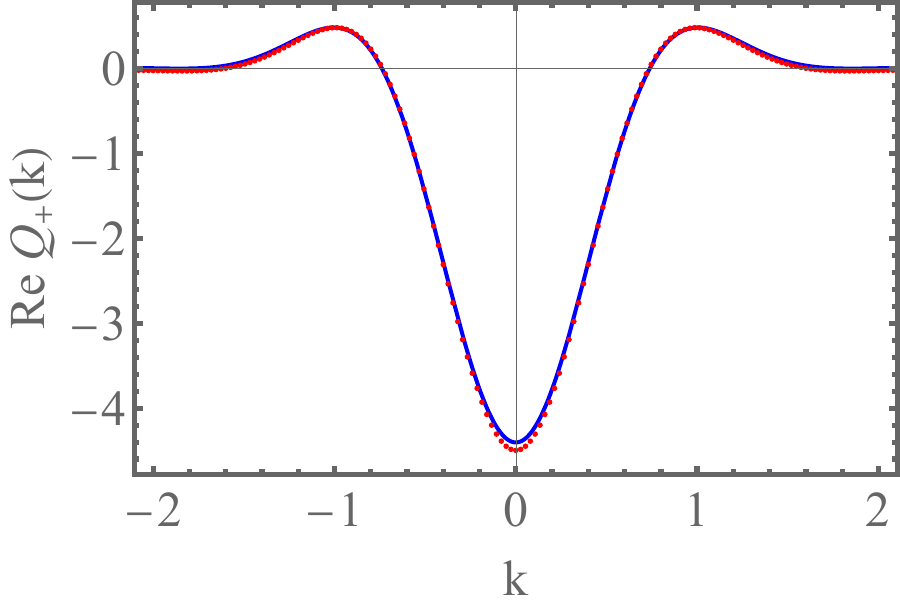}
\hspace{2mm}
\includegraphics[width=0.4\textwidth,clip=]{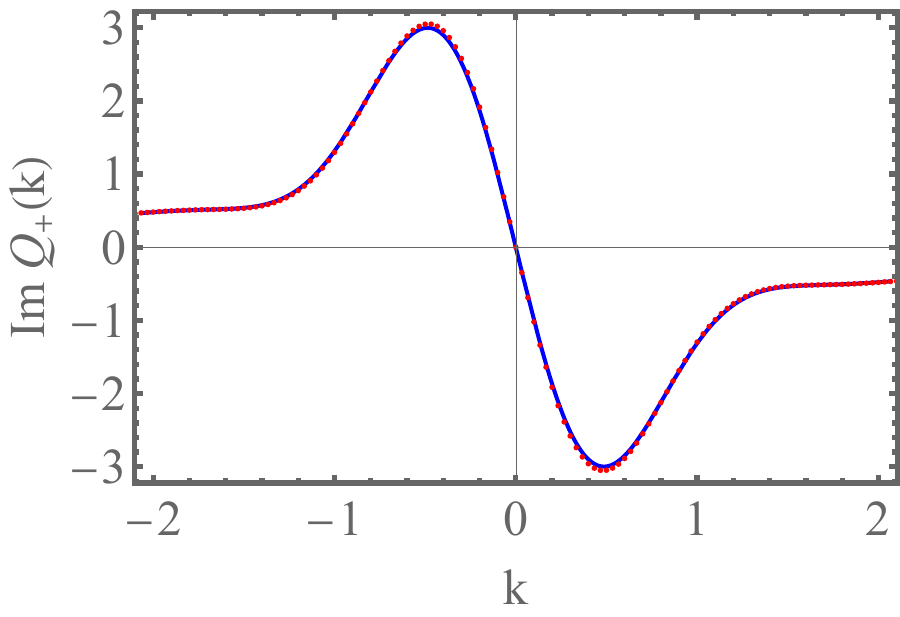}
\caption{Analytical results for the real and imaginary parts of $Q_+(k)$ for $X > \sqrt{8}$ and $\lambda > \lambda_{c2}$, described by Eqs. (\ref{adefPhi}), (\ref{Phi}) and  (\ref{equationwecompare})   (solid lines),
versus numerical results (markers). Parameters are $\lambda=20$ and $X=3.5$, corresponding to $J=0.70\dots$.}
\label{Qplus}
\end{center}
\end{figure}

Let us assume for now\be
\label{lambdaAssumption}
\lambda_{c2}< \min_{n=2,4,6,\dots}\lambda_{c}\left(X,n\right) \, ,
\ee
as this makes the analysis a little simpler.
At $\lambda=\lambda_{c0}$ two branch cuts cross the real axis. The branch points are denoted by $\kappa_1$, $\kappa_2$ and the points at which the branch cuts cross the real axis are denoted by $\pm k_{R0}$. The branch points $\kappa_1$ and $\kappa_2$, which satisfy $1+i\lambda\kappa_{i}e^{i\kappa_{i}X-\kappa_{i}^{2}}=0$, are reflections of each other with respect to the imaginary axis, $\kappa_1=-\kappa_2^*$. As $\lambda$ increases beyond $\lambda_{c0},$ rather than changing the contour of integration, one could instead represent the solution as

\begin{align}
\hat{a}^{(1)}_{\pm}\left(k\right)=\left(\frac{k_R^2}{\kappa_1\kappa_2}\frac{(k-\kappa_{1})(k-\kappa_{2})}{k_R^2-k^2}\right)^{\pm1}a_{\pm}\left(k\right),
\end{align}
while keeping the contour of integration in the definition of $a_\pm$, Eq. (\ref{adefPhi}), through $\Phi,$  Eq. (\ref{Phi}), to be along the real axis.
$\hat{a}^{(1)}_{\pm}\left(k\right)$ is a legitimate solution to Eq.~\eqref{apmEq}, as it is reached through a repeated use of the transformation described in Eq. (\ref{aHat})%
\footnote{One can take $k_R$ to have an infinitesimally small negative imaginary part, in order to keep $\hat{a}^{(1)}_{\pm}\left(k\right)$ analytic in the upper half complex $k$ plane.}.
It can be easily shown that that for $\lambda<\lambda_{c2}$ this is equivalent to just changing the contour of integration defining $a_\pm$ to avoid the branch cuts, while for $\lambda>\lambda_{c2}$ it is impossible to  find such a contour. This allows us to define for $\lambda_{c1}<\lambda<\lambda_{c2}$ two more solutions:
\begin{align}
&\hat{a}^{(2)}_{\pm}\left(k\right)=\left(\frac{k_R^2}
{\kappa_3\kappa_2}\frac{(k-\kappa_{3})(k-\kappa_{2})}
{k_R^2-k^2}\right)^{\pm1}a_{\pm}\left(k\right),\\[1mm]&
\hat{a}^{(3)}_{\pm}\left(k\right)=\left(\frac{k_R^2}
{\kappa_3\kappa_1}\frac{(k-\kappa_{3})(k-\kappa_1)}
{k_R^2-k^2}\right)^{\pm1}a_{\pm}\left(k\right).
\label{equationwecompare}\end{align}
Symmetry to reflection with respect to the imaginary axis, which is present since all $\kappa_i$ are purely imaginary at $\lambda_{c1} < \lambda < \lambda_{c2}$,  ensures that all three solutions produce real physical fields $u(x,1)$ and $v(x,0)$. Only $\hat{a}^{(1)}$ respects this symmetry for $\lambda<\lambda_{c1}$, where $\kappa_1$ and $\kappa_2$ are mirror images of each other. When $\lambda$ increases beyond $\lambda_{c2}$ the branch point $\kappa_2$ remains imaginary, while  $\kappa_1$ and $\kappa_3$ are mirror images of each other. This means that beyond $\lambda_{c2}$ only $\hat{a}^{(3)}$ yields real physical fields.
Figure  \ref{Qplus} shows $\text{Re}\, Q_+(k)$ and $\text{Im}\,Q_+(k)$ versus $k$ at $X=3.5$, $\lambda=20>\lambda_{c2}$, obtained by substituting Eqs. (\ref{adefPhi}) and (\ref{Phi}) in  Eq. (\ref{equationwecompare}).
This figure also shows the same quantities computed by solving  Eqs.~(\ref{d1}) and (\ref{d2}) numerically with a back-and-forth iteration algorithm \cite{CS}. The analytical and numerical curves are in good agreement.

\begin{figure}[ht]
 \includegraphics[width=0.3\textwidth,clip=]{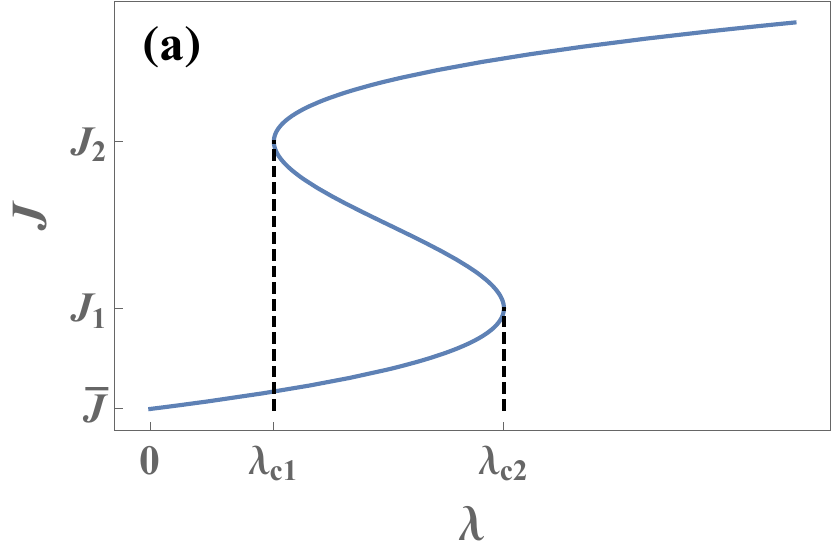}
 \includegraphics[width=0.3\textwidth,clip=]{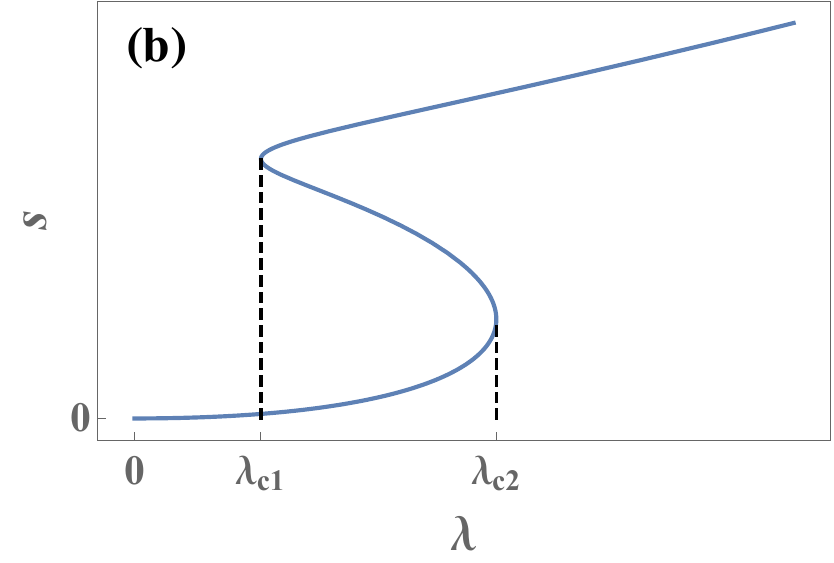}
\includegraphics[width=0.3\textwidth,clip=]{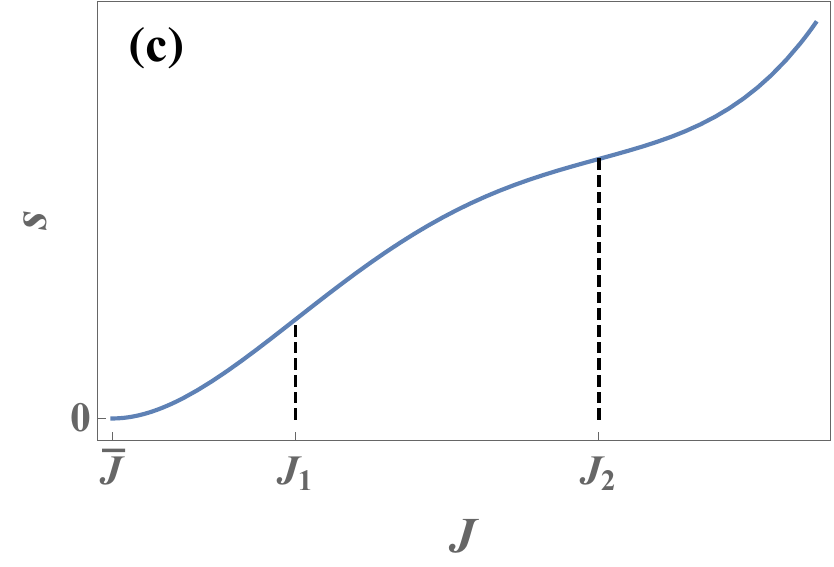}
\caption{Schematic plots of (a) $J(\lambda,X)$, (b) $s(\lambda,X)$ and (c) $s(J,X)$ for $X > \sqrt{8}$ and $\lambda > 0$. Interestingly, $J$ and $s$ are not single-valued functions of $\lambda$, and $s$ is not a convex function of $J$.}
\label{fig:sJlambdaSchematic}
\end{figure}

To avoid having to write additional terms in the equation for $a_\pm$ each time additional branch cuts cross the real axis, it is now convenient, after establishing which solutions are real, to change the contour of integration such that it does not cross any of these other branch cuts.
One may choose the  curve that satisfies  ${\rm Re}  \lambda ke^{ i  kX-k^{2}}=0$ and passes through the point $i\frac{X+\sqrt{X^2-8}}{4} $, see Figure \ref{figContourX} (b). For $\lambda>\lambda_{c2},$ this curve contains the branch points $\kappa_1$ and $\kappa_3$, and in computing the integral, the branch cut of the integrand must be drawn so as not to cross this curve. This change of integration contour from the real axis to this contour affects the change:
\begin{align}
a_\pm\to \left(\frac{k_R^2}{\kappa_1\kappa_2}\frac{(k-\kappa_1)(k-\kappa_2)}{k_R^2-k^2}\right)^{\pm 1} a_\pm,
\end{align}
for all $\lambda>\lambda_{c0}$, and leaves $a_\pm$ unchanged for $0<\lambda<\lambda_{c0}$. In this new representation (namely, for this choice of contour of integration), we have the following formulas (note that there are three branches of the solution at $\lambda_{c1}<\lambda<\lambda_{c2}$, whose origin will be explained shortly):
\be
\label{apmj1j2}
a_{\pm}\left(k\right)=\begin{cases}
e^{\Phi_{\pm}(k)-\Phi_{\pm}(0)}, & \lambda<\lambda_{c2},\\[2mm]
\left(\frac{\kappa_1}{\kappa_3}\frac{k-\kappa_{3}}{k-\kappa_{1}}\right)^{\pm1}e^{\Phi_{\pm}(k)-\Phi_{\pm}(0)}, & \lambda_{c1}<\lambda<\lambda_{c2},\\[3mm]
\left(\frac{\kappa_2}{\kappa_3}\frac{k-\kappa_{3}}{k-\kappa_{2}}\right)^{\pm1}e^{\Phi_{\pm}(k)-\Phi_{\pm}(0)}, & \lambda>\lambda_{c1}.
\end{cases}
\ee
As $J$ is increased, $\lambda$ grows up to the value $\lambda_{c2}$ along the branch corresponding to the first line in \eqref{apmj1j2}, up to a value $J=J_1$. As $J$ grows above $J_1$, $\lambda$ decreases down to a value $\lambda_{c1}$ which corresponds to some $J=J_2>J_1$. As $J$ is increased even further, $\lambda$ increases again: from $\lambda_{c1}$ to arbitrary large $\lambda$, so that $\lambda \to \infty$ corresponds to $J\to1$. The ensuing non-monotonic dependence of $\lambda$ on $J$ implies that $s$ is a non-convex function of $J$ (this is easy to see from $\partial_{J}s=\lambda$), which is quite an unusual situation in large-deviation theory. $J_1$ and $J_2$ are the inflection points, at which $\partial_{J}^{2}s=0$. These behaviors are schematically plotted in Fig.~\ref{fig:sJlambdaSchematic}.
We emphasize that Eq.~\eqref{apmj1j2} holds independently of the assumption \eqref{lambdaAssumption}, so it describes the solution at all $X>\sqrt{8}$ and at all $\lambda>0$.

We now calculate $Q_\pm(0)$, and from it we calculate $J=J(\lambda,X)$ and $s=s(\lambda,X)$. By following a similar route to the one above for $X<\sqrt{8}$ , we obtain  for $J\left(\lambda,X\right)=1 + Q_{+}\left(0\right)/\lambda$ 
the following expression:%
\footnote{In Eq.~\eqref{jcontinued} the term $1+$ from Eq.~\eqref{jlambda} is absent. The reason for this is that the integral in \eqref{jcontinued} is a contour integral in the upper half plane, see Fig.~\ref{figContourX} (b). When deforming the contour of integration from the real axis, as in \eqref{jlambda}, to the upper half plane, one must cross the pole of the integrand at $k=0$. The residue is $-1$, which exactly cancels the $+1$ term.}
\be
\label{jcontinued}
J\left(\lambda,X\right)= \int\frac{\ln\left(1+ i \lambda ke^{ i  kX-k^{2}}\right)}{k^{2}}\frac{dk}{2\pi\lambda}+\begin{cases}
0, & J<J_{1},\\[2mm]
\frac{ i }{\lambda}\left(\frac{1}{\kappa_{1}}-\frac{1}{\kappa_{3}}\right), & J_{1}<J<J_{2},\\[3mm]
\frac{ i }{\lambda}\left(\frac{1}{\kappa_{2}}-\frac{1}{\kappa_{3}}\right), & J>J_{2},
\end{cases}
\ee
and by integrating again  Eq.~\eqref{eq:dsdlambda} with  $s(\lambda=0,X)=0$, we find for the action:
\be
\label{scontinued}
\! s\left(\lambda,X\right)= \! \int \! \ln\left(1+ i \lambda ke^{ i  kX-k^{2}}\right)\frac{dk}{2\pi k^2}+ \! \int \! \text{Li}_{2}\left(-i k \lambda e^{ikX-k^{2}}\right)\frac{dk}{2\pi k^2} + \begin{cases}
0, & J<J_{1},\\[2mm]
2i\left(\kappa_{3}-\kappa_{1}\right)+X\ln\left(\frac{\kappa_{3}}{\kappa_{1}}\right), & J_{1}<J<J_{2},\\[3mm]
2i\left(\kappa_{3}-\kappa_{2}\right)+X\ln\left(\frac{\kappa_{3}}{\kappa_{2}}\right), & J>J_{2},
\end{cases}
\ee
a function which we plot in  Fig. \ref{svsj} as a function of $J$  for fixed $X=3$.
Note that the integration contour here is similar to the one drawn in Fig.~\ref{figContourX} (b). As such it is away from the real axis, where the integrands have a pole, and therefore no principal value integral is necessary.  Consequently, the Sokhotski--Plemelj formula is also not employed here in contrast, e. g.,  to Eq. (\ref{slambdasimpler}). To arrive at Eq.~(\ref{scontinued}), one has to evaluate the integral
\begin{align}
\int ^{\lambda}\frac{d\ln \lambda'}{\kappa_i(\lambda')} =\frac{\ln(\lambda)}{\kappa_{i}(\lambda)}+\int^{\kappa_i(\lambda)} \frac{\ln\lambda(\kappa_i')}{\kappa_i'^2}  d\kappa_i'.\label{howto}
\end{align}
By definition the root $\kappa'_i$ satisfies the following equation:
\begin{align}
\ln(\lambda(\kappa'_i))=\frac{ i  \pi}{2} -\ln\kappa'_i- i \kappa'_iX+\kappa'^2_i,
\end{align}
so that the integral on the right hand side of  Eq. (\ref{howto}) can be easily performed:
\begin{align}
&\int^{\lambda}\frac{ d\ln \lambda' }{\kappa_i(\lambda')} =- i  X\ln \kappa_i(\lambda) +\frac{1}{\kappa_i(\lambda)} +2 \kappa_i(\lambda)
\end{align}
allowing us to write Eq. (\ref{scontinued}).

\section{Asymptotic limits}
\label{asymptot}

Our exact results greatly simplify in three asymptotic limits that we present here. Two of the three asymptotic formulas can be obtained either from the exact results, or directly from the MFT equations by using two different perturbation methods.

\subsection{$\lambda \ll 1$: linear theory}

In the limit of small $\lambda$ the optimal path can be found by a straightforward perturbative expansion around
the zeroth-order solution (\ref{meanfield}) \cite{KrMe}.  Here, due to the symmetry relation (\ref{uvsymmetry}), the leading order calculation in $\lambda \ll 1$ is especially simple \cite{BSM1}. Using Eq.~(\ref{action0}), we obtain
\be
\label{actionsimple}
  s(\lambda,X)\simeq \lambda^2 \int_0^1  dt \int_{-\infty}^{\infty}  dx\, \bar{u}^2(x,t)\,\bar{u}^2(X-x,1-t)
  = \frac{\lambda ^2 e^{-\frac{X^2}{2}}}{8
   \sqrt{2 \pi }}\,,
\ee
where we have plugged the zeroth-order solution (\ref{meanfield}), rescaled by $W$. The same result~(\ref{actionsimple}) follows from the exact rate function, described by Eqs.~(\ref{jlambda}) and (\ref{slambdasimpler}), via a straightforward expansion in $\lambda$.

Rewriting the shortcut relation $\partial s/\partial J = \lambda$ in the form of $(\partial s/\partial\lambda) (\partial\lambda/\partial J) = \lambda$, and combining it with Eq.~(\ref{actionsimple}), we can express $\lambda$ in terms of $J$ and $X$:
\begin{equation}\label{lambdalinear}
\lambda\simeq 4 \sqrt{2\pi} e^{\frac{X^2}{2}} (J-\bar{J})\,.
\end{equation}
Equations~(\ref{actionsimple}) and (\ref{lambdalinear}) yield (still in the rescaled variables) a quadratic in $J-\bar{J}$ asymptotic which strongly depends on $X$:
\begin{equation}\label{slinear}
s(J\to \bar{J},X)\simeq 2 \sqrt{2\pi}\, e^{\frac{X^2}{2}} (J-\bar{J})^2\,.
\end{equation}
The corresponding probability distribution $\mathcal{P}(J,X,T)$ describes typical, small Gaussian fluctuations of the transferred heat around its mean value $\bar{J}$. In the original variables
\begin{equation}\label{probsmall}
\ln \mathcal{P}_{\text{Gauss}} \simeq - 2\sqrt{2\pi T}\,e^{\frac{X^2}{2T}}\left[\frac{J}{W}-\frac{1}{2} \text{erfc}\left(\frac{X}{\sqrt{4T}}\right)\right]^2\,.
\end{equation}
In the particular case $X=0$ this Gaussian asymptotic coincides with the one obtained in Ref. \cite{BSM1}.  Note that Eq.~(\ref{slinear}) is applicable only for  $|\lambda|\ll 1$, that is for $|J-\bar{J}|\ll e^{-X^2/2}$. This condition becomes restrictive at $X\gg 1$, where the expected value
of the transferred heat is extremely small.

Evaluating Eqs.~(\ref{lambdalinear}) and (\ref{slinear}) at $J=1/2$, one can probe small fluctuations of the energy median $x=X_m$, which is an analog of the tracer position in the SSEP and other single-file systems.  At fixed $J$ the applicability condition of the perturbation expansion $\lambda\ll 1$ demands that $|X_m|\ll 1$. Using the small-argument asymptotic of $\text{erfc}(...)$, we obtain in the leading order $\lambda \simeq 2\sqrt{2} X_m$, and (in the rescaled variables)
\begin{equation}\label{stracerlin}
s\simeq \frac{X_m^2}{\sqrt{2\pi}}\,.
\end{equation}
This corresponds to a Gaussian asymptotic of the probability distribution $\rho(X_m,T)$ of the median. In the original variables
\begin{equation}\label{typicalX}
\ln \rho(X_m,T) \simeq - \frac{X_m^2}{\sqrt{2\pi T}}\,.
\end{equation}
The standard deviation of the energy median position, $\langle X_m^2 \rangle^{1/2} = (\pi/2)^{1/4} T^{1/4}$, exhibits the familiar one-dimensional tracer scaling with time, $T^{1/4}$.

\subsection{$\lambda \gg 1$, $X\gg 1$}

\subsubsection{From adiabatic perturbation theory}

The large-deviation limit $\lambda \gg 1$ and $X\gg 1$ is especially interesting. The reason for that is a unique
way of the most probable transfer of a significant amount of energy to large distances,  specific for lattice-gas models  (such as the KMP) for which the second derivative of the compressibility $\sigma(u)$ with respect to $u$ \cite{MS2013} is positive. In this case the optimal path has the form of a \emph{traveling doublon}: a pair of coupled large-amplitude solitary pulses of $u$ and $v$  \cite{MS2013,ZarfatyM}. The amplitudes of the $u$ and $v$ pulses slowly (adiabatically) vary with time  \cite{MS2013}. An adiabatic perturbation theory, which describes this nontrivial process in our present setting, is a slightly modified version of the perturbation theory of Ref. \cite{MS2013}, where a step-like initial $u$-profile were considered, and $X$ was zero.

The adiabatic perturbation theory of Ref. \cite{MS2013} exploits the strong inequality $\lambda\gg 1$ which guarantees a large length-scale separation between the relatively small \emph{interior} region of the doublon and the large-scale \emph{exterior} regions in front of the doublon and behind it. In the exterior regions one can neglect the diffusion terms in the MFT equations (\ref{d1}) and (\ref{d2}).  The diffusion, however, is important for the description of the doublon itself (the interior region).  Importantly, when viewed from the exterior regions, the doublon can be effectively described as a single \emph{point-like singularity} located at a point $x = X_d(t)$ which travels in space. The
singularity carries a finite energy $M(t)$, a finite jump of the ``momentum" $V(t) = p_1 -  p_2$ (where the subscripts 1 and 2
refer to ``in front of the singularity" and ``behind the singularity", respectively), and a finite value of the Hamiltonian  $\mathcal{H}(t)$, corresponding to the MFT equations (\ref{d1}) and (\ref{d2}) rewritten in the Hamiltonian form (that is, in the canonical variables $u$ and $p$). The quantities $M(t)$, $V(t)$ and $\mathcal{H}(t)$ slowly depend on time. They are formally defined as
\begin{equation}\label{PM060}
    \{M(t),\,V(t),\,{\mathcal H}(t)\} = \int_{X_d(t)-\varepsilon}^{X_d(t)+\varepsilon} \{u,\,-v,\,h\}\, dx,
\end{equation}
where $h=v\,\partial_x u +u^2 v^2$ is the density of the Hamiltonian of Eqs. (\ref{d1}) and (\ref{d2}) rewritten in terms of the canonical variables $u$ and $p$, and $\varepsilon>0$ can be fixed at an arbitrary small value as $X\to\infty$ and $\lambda\to \infty$.  It was shown by direct calculations in Ref. \cite{MS2013} that
\begin{equation}\label{relations}
M(t)\simeq \frac{1}{\tilde{v}_1(t)}, \quad V(t)\simeq \frac{1}{u_2(t)} ,  \quad \text{and}\quad \mathcal{H}(t)\simeq c(t),
\end{equation}
where $\tilde{v}_1(t)$ is the value of $v$ in front of the doublon, and $u_2(t)$ is the value of $u$ behind the doublon.
As we will see shortly, the doublon velocity $c(t)$ is actually constant. Using  Eqs. (57)-(59) and (62)-(64) of Ref. \cite{MS2013} and the relation $\dot{X}_d (t) = c(t)$, we arrive at the equations
\begin{eqnarray}
\dot{M} &=& -c u_2,
\label{PM110}\\
\dot{V} &=& c \tilde{v}_1,
\label{PM112}\\
\dot{{\cal H}} &=& 0.
\label{PM114}
\end{eqnarray}
Then, using Eqs.~(\ref{relations}), we obtain
\begin{eqnarray}
\dot{M} &=& -\frac{c}{V}\,,
\label{PM115}\\
\dot{V} &=& \frac{c}{M}\,,
\label{PM116}\\
c&=&\text{const}\,.
\label{PM117a}
\end{eqnarray}
The doublon must arrive at $t=1$ at the point $x=X$, which gives $c=X$. Equations (\ref{PM115}) and (\ref{PM116}) have the first integral
\begin{equation}\label{firstint}
M(t)V(t)=k_1(X)\,,
\end{equation}
where the function $k_1(X)$ is a priori unknown. Using this first integral in Eq.~(\ref{PM115}), we obtain
\begin{equation}\label{eqM}
M(t) = \exp\left[k_2(X)-\frac{Xt}{k_1(X)}\right]\,.
\end{equation}
The a priori unknown function $k_2(X)$ must be set to zero in view  of the condition $M(t=0)=1$, and we obtain
 \begin{equation}\label{PM117}
M(t) = \exp\left[-\frac{Xt}{k_1(X)}\right]\,.
 \end{equation}
Then Eq.~(\ref{firstint}) yields
 \begin{equation}\label{PM118}
V(t) = k_1(X) \exp\left[\frac{Xt}{k_1(X)}\right]\,.
 \end{equation}
Using the two remaining boundary conditions \cite{smallboundarylayers},
\begin{equation}\label{2bound}
M(t=1) = j \quad \text{and} \quad V(t=1) = \lambda\,,
\end{equation}
we determine $k_1(X)$ and $\lambda$:
\begin{equation}\label{largeresults}
k_1(X) = \frac{X}{\ln \frac{1}{j}} \quad \text{and} \quad \lambda = \frac{X}{j \ln \frac{1}{j}}\,.
\end{equation}
Using the first of these relations in Eq.~(\ref{PM117}), we see that the energy $M(t)$ of the doublon decays exponentially in time with rate $\ln (1/j)$:
\begin{equation}\label{Mdecays}
M(t) = \exp\left(-t\ln \frac{1}{j}\right)\,.
\end{equation}
In its turn, the doublon's momentum $V(t)$
grows exponentially in time with the same rate:
\begin{equation}\label{Vgrows}
V(t) = \frac{X}{\ln \frac{1}{j}}\exp\left(t\ln \frac{1}{j}\right)\,,
\end{equation}
which completes the task of determining the optimal path of the doublon in the adiabatic approximation.
Note that $M(t)$ and $V(t)$ from Eqs.~\eqref{Mdecays} and \eqref{Vgrows} respectively, together with
$\lambda$ from Eq.~\eqref{largeresults},
satisfy the symmetry relation \eqref{uvsymmetry} integrated over the doublon region
$(X_d(t)-\varepsilon, X_d(t)+\varepsilon)$.

The rescaled action, according to Ref. \cite{MS2013}, is
\begin{equation}\label{actionlarge}
s(j,X) \simeq c \ln (c M V) =  X \ln \frac{X^2}{\ln \frac{1}{j}}\,, \quad X\gg 1.
\end{equation}
The corresponding large-$X$ asymptotic of the probability distribution is
\begin{equation}\label{largeprob}
\ln \mathcal{P}(J,X,T) \simeq  - X \ln \left(\frac{X^2}{T \ln \frac{W}{J}}\right)\,,\quad X\gg \sqrt{T}\,.
\end{equation}
This expression describes a faster-than-exponential tail of $\mathcal{P}$ as a function of $X$. This tail depends
only logarithmically on $T$, and extremely weakly on $J/W$.

\subsubsection{From exact rate function}

For large $X$ we are automatically in the regime of $X>\sqrt{8}$. As the contour of integration we can use the curve $\arg i k e^{ i kX -k^2}=0$, which at ${\rm Re} k\to\pm\infty $ merges with the line ${\rm Im} k=\frac{X}{2}$. In fact, one can ascertain that for large $X$ the entire curve $\arg i k e^{ i kX -k^2}=0$ has ${\rm Im} k\simeq\frac{X}{2}$. Then, in view of  Eq. (\ref{scontinued})  for the action, it can be shown that  the integrand is vanishingly small unless $\lambda\sim e^{\frac{X^2}{4}}$. In addition, we shall see a posteriori that we obtain all possible values of $J=O(1)$ by keeping $\lambda$ much smaller than $e^{\frac{X^2}{4}}$. In fact, we obtain $J_1=0$ and $J_2=e^{-1}$.
Given all this, the action in this region can be written as:
\begin{align}\label{SlargeX}
s\left(\lambda,X\right)\simeq \begin{cases}
2i\left(\kappa_{3}(\lambda)-\kappa_{1}(\lambda)\right)+X\ln\left(\frac{\kappa_{3}(\lambda)}{\kappa_{1}(\lambda)}\right), & J_{1}<J<J_{2},\\[3mm]
2i\left(\kappa_{3}(\lambda)-\kappa_{2}(\lambda)\right)+X\ln\left(\frac{\kappa_{3}(\lambda)}{\kappa_{2}(\lambda)}\right), & J>J_{2}\,.
\end{cases}
\end{align}
The expression for the current, Eq. (\ref{jcontinued}), also benefits from the vanishing of the integral, and  we can write approximately:
\be
\label{jLargeX}
J\left(\lambda,X\right)\simeq\begin{cases}
\frac{ i }{\lambda}\left(\frac{1}{\kappa_{1}}-\frac{1}{\kappa_{3}}\right), & J_{1}<J<J_{2},\\[3mm]
\frac{ i }{\lambda}\left(\frac{1}{\kappa_{2}}-\frac{1}{\kappa_{3}}\right), & J>J_{2}\,.
\end{cases}\ee

As we shall also see, for $J=O(1)$, that  $\lambda\simeq X$ and $\kappa_{1,2}\sim \frac{1}{ \lambda}$. Therefore we can approximate Eq. (\ref{kiEq})  for $\kappa_j,$ $j\in \{1,2\}$  as follows:
\begin{align}
- \frac{1}{\lambda}= i \kappa_j e^{ i\kappa_jX} .
\end{align}
This algebraic equation for $\kappa_j$ can be solved in terms of the Lambert function $W_{j-2}$:
\begin{align}
\kappa_j=-\frac{i  }{X}W_{j-2}\left(-\frac{X}{\lambda}\right)\,,\quad j\in\{1,2\} \,.
\end{align} As for $\kappa_3,$ in the large $X$ limit, where $\lambda\sim X,$ one finds that $\kappa_3$ is of order $X$\ and has the following form:
\begin{align}
\kappa_3= i X- i \frac{\ln(\lambda X)}{X}.
\end{align}Then $J$\ approximates to\begin{align}
J\sim \frac{ i}{\lambda  \kappa_j}=-\frac{X}{\lambda}W_{j-2}^{-1}\left(-\frac{X}{\lambda}\right), \quad j\in\{1,2\}
\end{align}
Since this equation can give any $J$ from $0$ to $1$, we thus confirm that $J_1=0$, and only the second and third branches are relevant in Eqs. (\ref{jcontinued}) and (\ref{scontinued}).
We also see that $J_2$ is obtained at the branch  point of the Lambert function, $-e^{-1}$. Namely $J_2=\frac{-e^{-1}}{W_0(-e^{-1})} = e^{-1}$.

To find $s(J)$  from Eq.~(\ref{SlargeX}) for $s(\lambda)$, we first write $J=\frac{i}{\lambda  \kappa_j}=e^{ i \kappa_i X},$ the last equality being a consequence of the definition of $\kappa_i.$ This gives $\kappa_i=- i  \frac{\ln J}{X}$. Then, using again the equality $J=\frac{ i}{\lambda  \kappa_j}$, we obtain the desired relation for $\lambda$ in terms of $X$ and $J$,
\begin{align}
  \quad \lambda=\frac{X }{J\ln \frac{1}{J}}\,,
\end{align}
which coincides with our second perturbative relation in Eq. (\ref{largeresults}).
Using this expression, we can rewrite $\kappa_j$ in terms of $J$\ only:
\begin{align}
\kappa_{j}=\begin{cases}
\frac{i\ln(1/J)}{X}, & j\in\{1,2\},\\[3mm]
iX-\frac{i}{X}\ln\left(\frac{X^{2}}{J\ln(1/J)}\right), & j=3,
\end{cases}
\end{align}

Then one can substitute this into the action Eq. (\ref{SlargeX}) and arrive at
\begin{equation}
s(J,X) =X\ln \left(\frac{X^2}{e^2\ln \frac{1}{J}}\right) +O(1)\,.
\label{actionlargeexact}
\end{equation}
A comparison of Eq.~(\ref{actionlargeexact}) with the perturbative result (\ref{actionlarge}) shows that the latter misses an important subleading factor $e^{-2}$ under the external logarithm.

\subsection{$\lambda\gg 1$ and $X$ held constant: $J\to 1$}

For large $\lambda$ we are automatically on branch $3$. We assume here that $X>\sqrt{8},$ but the same result can be obtained for $X<\sqrt{8}$ by largely the same means, so we do not expound on the case $X<\sqrt{8}$ here.          The integration contour for the integrals in question can be performed on the curve  $\arg i  k e^{ i  kX -k^2}=0$, which at ${\rm Re} \, k\to\pm\infty $ merges with the line ${\rm Im}\,k=X/2,$ as shown in Fig.~\ref{figContourX} (b).

We shall need the asymptotic expression for $\kappa_i$ in this limit. These are easily computed from Eq. (\ref{kiEq}) to be
\begin{align}
&\kappa_2^{-1}=- i  \lambda+ i  X-\frac{ i }{  \lambda}+O\left(\frac{1}{\lambda^2}\right)\\
&\kappa_3=-\kappa_1^*=\frac{ i  X}{2} +\sqrt{\ln \lambda}-\frac{ i  \pi+\frac{X^2}{2}-\ln \ln \lambda}{4 \sqrt{\ln \lambda }}+O\left(\frac{1}{\ln\lambda}\right).
\end{align}

Computing the first integral in Eq. (\ref{scontinued}), we may first note the symmetry $k\to-k^*, $  under which the integral simply gets complex conjugated. This means that one may first compute only the integral on that  part of the contour  in Fig. \ref{figContourX} (b) that runs from $\infty$ to the imaginary axis, and then take the real value. We decompose this contour into three parts, the first runs from $\infty $ to $\kappa_3+\delta A, $ where $\delta A\sim (\ln\lambda)^{-1/4}$. The integrand in this region is vanishingly small and can be dropped. The second part of the contour runs on a small circular arc of radius $\delta A$ and center $\kappa_3$ runnig from $\kappa_3+\delta A$ to $\kappa_3-\delta A$.  On this part of the contour the expression inside the logarithm can be expanded around $\kappa_3.$  One gets:\begin{align}
&\int \frac{\ln\left(1+  i  \lambda ke^{ i  kX-k^2}\right)}{k^2} \frac{dk}{2\pi  }=\nonumber \frac{\delta A^2}{2\pi  \kappa_3}+O\left(\frac{1}{(\ln\lambda)^{3/2}}\right).
\end{align}
For the third part of the contour,  we define $A= \kappa_3+\delta A$ and integrate over that part of the contour  running from $A$ to the imaginary axis along the contour of the right panel of Fig. \ref{figContourX}.
In fact, it is easier to integrate from $A$ to $-A^*,$ which is ultimately what is needed. Given that, we write:
\begin{align}
\int^{A}_{-A^*} \frac{\ln\left(1+  i  \lambda ke^{ i  kX-k^2}\right)}{k^2} \frac{dk}{2\pi  }&=\int^{A}_{-A^*} \frac{\ln\left(  i  \lambda k\right) + i  kX-k^2}{k^2} \frac{dk}{2\pi  }+O\left(\frac{1}{\lambda}\right)\nn\\
&=-\left.\frac{\ln\left(  i  \lambda k\right)+1 }{2\pi k}+\frac{ i  X \ln k -k}{2\pi }\right|_{-A^*}^A +O\left(\frac{1}{\lambda}\right) \, .
\end{align}
We now expand the last expression taken at the upper limit near $\kappa_3$ and use
\be
\label{k3Property}
\ln( i   \lambda \kappa_3)=- i  \pi- i  \kappa_3X+\kappa_3^2
\ee
to simplify the combined results of the calculation. Applying the symmetry $k\to-k^*, $   we find
for $J$,  which in turn is given by Eq. (\ref{jcontinued}), the following:
\begin{align}
J(\lambda, X)&=
\int \frac{\ln\left(1+  i  \lambda ke^{ i  kX-k^2}\right)}{\lambda k^2} \frac{dk}{2\pi  }-\frac{ i  }{\lambda \kappa_{3}}+\frac{ i  }{\lambda \kappa_{2}} \nn\\
&\simeq\frac{Q_+(0)}{\lambda} \simeq1-\frac{2(\ln \lambda )^{1/2}}{\pi \lambda }-\frac{(\ln \lambda )^{-1/2}}{\pi\lambda }\left(1+\frac{X^2}{4} +\frac{1}{2}\ln \ln\lambda\right),
\end{align}
from which one gets:
\begin{align}
-\frac{\pi^{2}}{2}(1-J)^{2}=- \frac{2\ln\lambda}{\lambda^2} \left[1+\frac{1+\frac{X^{2}}{4}+\frac{1}{2}\ln\ln\lambda}{\ln\lambda}
+O\left(\frac{1}{\ln^{2}\lambda}\right)\right]
\end{align}
Thus to this order:
\begin{align}
\ln\lambda = -\frac{1}{2}W_{-1}\left(-\frac{\pi^2}{2}(1-J)^2\right)+1+\frac{X^2}{4} +\frac{1}{2}\ln \left[-\frac{1}{2}W_{-1}\left(-\frac{\pi^2}{2}(1-J)^2\right)\right].\label{lambdaofJasympt}
\end{align}

To compute the action we use again Eq. (\ref{eq:dsdlambda}) to write:
\begin{align}
s(\lambda, X)=Q_+(0)-\int Q_+(0)\frac{d\lambda}{\lambda}\simeq\frac{4(\ln \lambda )^{3/2}}{3\pi  }+\frac{{X^2} }{\pi } (\ln \lambda )^{1/2} \, .
\end{align}
Re-writing this in terms of $J$ instead of $\lambda$ using Eq. (\ref{lambdaofJasympt}), we finally get, in the limit $J \to 1$ at fixed $X$,
\begin{align}
s(J\to 1,X)&\simeq\frac{ 4}{3\pi}\left[-\frac{1}{2}W_{-1}\left(-\frac{\pi^2}{2}(1-J)^2\right)\right]^{3/2}\nn\\
&+\frac{2}{\pi}\left[-\frac{1}{2}W_{-1}\left(-\frac{\pi^{2}}{2}(1-J)^{2}\right)\right]^{1/2}
\left[1+\frac{3X^{2}}{4}+\frac{1}{2}
\ln\left(-\frac{1}{2}W_{-1}\left(-\frac{\pi^{2}}{2}(1-J)^{2}\right)\right)\right] \, .
\label{twoterms}
\end{align}
The leading term, that is the first term on the right-hand side of  Eq.~(\ref{twoterms}), coincides with the result obtained in Ref. \cite{BSM1} for the particular case $X=0$. As one can see,
a finite $X$ affects the $J\to 1$ asymptotic only in a subleading order, described by the second term on the right-hand side of  Eq.~(\ref{twoterms}).

\section{Discussion}

\label{discussion}

Here we studied the full distribution $\mathcal{P}(J,X,T)$ of the total heat $J$ transferred to the right of a point $x=X$ in the KMP model at long times for an initially localized heat pulse, thereby extending our previous work \cite{BSM1} which was limited to $X=0$.
We calculated exactly the rescaled rate function $s(J,X)$ that describes this distribution, by combining the MFT and the ISM. In particular we found that for $X>\sqrt{8}$ (in properly rescaled variables) $s$ is not convex as a function of $J$, a rather unusual situation in large-deviation problems.
Furthermore, we studied  three different asymptotics of $s(J,X,T)$ by extracting them from the exact solution and also obtaining two of them by applying two different perturbation methods directly to the MFT equations. The perturbative solutions also yield the optimal histories of the system conditioned on $J$. This gives an important additional physical insight, at present unaccessible for the ISM.
The distribution $\mathcal{P}(J,X,T)$ as a function of $X$ at fixed $J$ has the same long-time dynamical scaling as the distribution of a tracer particle in a single-file transport of interacting particles, as in the SSEP.
It would be interesting to try to apply the ISM to other large-deviation problems for the KMP model, in which non-stationary heat transfer statistics problems were previously solved analytically only in some limiting cases \cite{DG2009b,KrMe,MS2013}.

The MFT of lattice gases can be viewed as a particular case of the optimal fluctuation method (OFM), a versatile tool which can be used to study large deviations in a broad class of macroscopic systems that also includes paradigmatic surface growth models (such as the KPZ equation), reaction-diffusion systems, and many more \cite{Halperin,Langer,Lifshitz, turb1, EK,MS2011, Fogedby,KK,MKV,HMS}. 
Mathematically, the OFM formulation involves a set of  coupled nonlinear partial differential equations whose solution gives the optimal history of the system, conditioned on the rare event in question. It is generally very hard to solve these equations exactly.
Therefore, uncovering and exploiting exact integrability of these equations appears to be a way to make a significant progress in some of these systems.
Indeed, the present work is one of several very recent studies \cite{BSM1, KLD1,KLD2, MMS22,KLD3} which use the ISM in order to obtain exact solutions for full statistics of different physical quantities in nonlinear spatio-temporal stochastic systems. We believe that this line of work has a great potential for the future.

\textit{Note added.} Ref. \cite{KLD3} was posted shortly before the present paper was submitted. The authors of Ref. \cite{KLD3} focused on a crossover between different large-deviation regimes for a model which is mathematically identical to the KMP model. It appears that their work is closely related to ours, and it will be interesting to make a detailed comparison of the results.

\begin{appendices}

\section{Derivation of the MFT equations and boundary conditions}
\label{app:MFT}
\renewcommand{\theequation}{A\arabic{equation}}
\setcounter{equation}{0}

Let us first explain why the effective amplitude of the noise becomes small in the long-time limit $T \gg 1$.
We rescale time and space, $\tilde{t}=t/T$, $\tilde{x}=x/\sqrt{T}$, and define
$\tilde{u}\left(\tilde{x},\tilde{t}\right)\equiv u\left(\tilde{x}\sqrt{T},\tilde{t}T\right)$,
and the dimensionless noise term
$\tilde{\eta}\left(\tilde{x},\tilde{t}\right)=T^{3/4}\eta\left(\tilde{x}\sqrt{T},\tilde{t}T\right)$.
Under these rescalings, Eq.~\eqref{Langevin} of the main text becomes
\be
\partial_{\tilde{t}}\tilde{u}=
\partial_{\tilde{x}}\left(\partial_{\tilde{x}}\tilde{u}+T^{-1/4}\sqrt{2}\,
\tilde{u}\tilde{\eta}\right)
\ee
from which it is clearly observed that the noise is effectively weak in the limit $T \gg 1$. Note, however, that the long-time limit must be taken with constant $X / \sqrt{T}$.

We now derive of the MFT equations, Eqs.~\eqref{d1} and \eqref{d2} of the main text.
We first define a ``potential''
\be
\psi\left(x,t\right)=\int_{-\infty}^{x}u\left(y,t\right)dy \, .
\ee
Integrating Eq.~\eqref{Langevin} of the main text with respect to $x$ and using the boundary condition $u(x\to-\infty,t)\to0$, we obtain a Langevin equation for $\psi(x,t)$:
\be
\label{psiLangevin}
\partial_{t}\psi=\partial_{x}^{2}\psi+\sqrt{2}\,\partial_{x}\psi\,\eta\,.
\ee
An equivalent description of the dynamics \eqref{psiLangevin} is provided by the path-integral representation. The probability density functional $P[\eta]$ of the white Gaussian noise term $\eta(x,t)$ is
\be
\label{Peta}
P\left[\eta\right]\sim\exp\left(-\int_{0}^{T}dt\int_{-\infty}^{\infty}dx\,\frac{\eta^{2}}{2}\right)\,.
\ee
The advantage of using the $\psi$ representation is that the noise term in Eq.~\eqref{psiLangevin} appears without a spatial derivative, and one can therefore express $\eta$ through $\psi$ via Eq.~\eqref{psiLangevin}.
Using this in Eq.~\eqref{Peta}, one finds that the probability density for a given realization of $\psi(x,t)$ is:
\be
\label{Spsi}
-\ln P\left[\psi\right]\simeq S\left[\psi\right]\equiv\frac{1}{4}\int_{0}^{T}dt\int_{-\infty}^{\infty}dx\,\left(\frac{\partial_{t}\psi
-\partial_{x}^{2}\psi}{\partial_{x}\psi}\right)^{2}\,,
\ee
where $S\left[\psi\right]$ is the action functional.
As explained above, in the limit $\sqrt{T}\gg 1$, the noise is effectively weak. This enables us to calculate $\mathcal{P}(J,T)$ via a saddle-point evaluation of the path integral. Within this framework, $-\ln\mathcal{P}\left(J,T\right)$ is given by the minimum of the action functional $S$ with respect to all realizations $\psi(x,t)$, constrained on the initial condition $u(x,0)=W\delta(x)$ and on a given value of the transferred heat
$J=W-\psi(X,T)$ (where we used the conservation law $\int_{-\infty}^{\infty} u(x,t=T) \,dx = W$).
The latter constraint on $J$ is taken into account by adding the term $-\Lambda\psi\left(0,T\right)$ to the action, where $\Lambda$ is a Lagrange multiplier, i.e., we minimize the constrained functional
\be
S_{\Lambda}\left[\psi\right]=S\left[\psi\right]-\Lambda\psi\left(0,T\right)=\frac{1}{4}\int_{0}^{T}dt\int_{-\infty}^{\infty}dx\,\left(\frac{\partial_{t}\psi-\partial_{x}^{2}\psi}{\partial_{x}\psi}\right)^{2}-\Lambda\psi\left(X,T\right) \, .
\ee

To minimize $S_{\Lambda}$, we must require its linear variation with respect to $\psi(x,t)$ to vanish. 
A given variation $\psi(x,t) \to \psi(x,t)+\delta \psi(x,t)$ causes a variation of $S_{\Lambda}\left[\psi\right]$ which, to first order in $\delta \psi$, is given by
\bea
\label{deltaS1}
\delta S_{\Lambda} &=&S_{\Lambda}\left[\psi+\delta\psi\right]-S_{\Lambda}\left[\psi\right]=\frac{1}{2}\int_{0}^{T}dt\int_{-\infty}^{\infty}dx\,\left(\frac{\partial_{x}^{2}\psi-\partial_{t}\psi}{\partial_{x}\psi}\right)\left[\frac{\partial_{x}^{2}\delta\psi-\partial_{t}\delta\psi}{\partial_{x}\psi}-\frac{\partial_{x}^{2}\psi-\partial_{t}\psi}{\left(\partial_{x}\psi\right)^{2}}\partial_{x}\delta\psi\right]-\Lambda\delta\psi\left(X,T\right)\nn\\
&=& \int_{0}^{T}dt\int_{-\infty}^{\infty}dx\,\partial_{x}p\left(\partial_{x}^{2}\delta\psi-\partial_{t}\delta\psi-2\partial_{x}p\partial_{x}\psi\partial_{x}\delta\psi\right)-\Lambda\delta\psi\left(X,T\right)\,,
\eea
where we have defined the momentum density gradient
\be
\label{pdef}
\partial_{x}p=\frac{\partial_{x}^{2}\psi-\partial_{t}\psi}{2\left(\partial_{x}\psi\right)^{2}} \; .
\ee
Integrating by parts in Eq.~\eqref{deltaS1}, we obtain
\bea
\label{deltaS2}
\delta S_{\Lambda}&=&\int_{0}^{T}dt\int_{-\infty}^{\infty}dx\,\left\{ \partial_{x}^{3}p+\partial_{xt}p+2\partial_{x}\left[\left(\partial_{x}p\right)^{2}\partial_{x}\psi\right]\right\} \delta\psi \nn\\
&+&\int_{-\infty}^{\infty}dx\,\left[\partial_{x}p\left(x,0\right)\delta\psi\left(x,0\right)-\partial_{x}p\left(x,T\right)\delta\psi\left(x,T\right)-\Lambda\delta\left(x - X\right)\delta\psi\left(x,T\right)\right]\,,
\eea
where the first two terms in the single integral are the boundary terms originating from the integration by parts in time.
Since $\psi(x,0)$ is given (i.e., the initial condition is quenched), the variation $\delta\psi\left(x,0\right)$ vanishes.

By requiring the double integral in \eqref{deltaS2} to vanish for arbitrary $\delta \psi$, one obtains the second MFT equation, Eq.~\eqref{d2} in the main text (recalling that $v=-\partial_{x}p$ and $u=\partial_{x} \psi$). The first MFT equation, Eq.~\eqref{d1} in the main text, follows from Eq.~\eqref{pdef} after multiplying by the denominator and then taking a spatial derivative.
Note that, under the rescalings of $x$, $t$ and $u$ described in the text, $v$ and $X$ should be rescaled by $1/W$ and $\sqrt{T}$, respectively. The MFT equations invariant under these rescalings.
In addition, we must require the single integral in Eq.~\eqref{deltaS2} to vanish for arbitrary $\delta\psi\left(x,T\right)$. This yields the boundary condition
\be
\partial_{x}p\left(x,T\right) = -\Lambda\delta(x - X) \, .
\ee
After the rescaling, this becomes Eq.~\eqref{vdelta} in the main text, where $\lambda = W\Lambda$ is a rescaled Lagrange multiplier, and $X$ is rescaled as well.
Finally, using Eq.~\eqref{pdef} in \eqref{Spsi} one finds that the action can be rewritten as $S=\int_{0}^{T}dt\int_{-\infty}^{\infty}dx\,u^{2}v^{2}$ which, after the rescaling, leads to $S=\sqrt{T} \,  s$ where the rescaled action $s$ is given by Eq.~\eqref{action0} in the main text.
The large parameter $S\sim\sqrt{T}\gg1$ justifies \textit{a posteriori} the saddle-point approximation that we used.

\section{Solving the scattering problem at $t=0$}
\label{app:t0}

\renewcommand{\theequation}{B\arabic{equation}}
\setcounter{equation}{0}

In this appendix, we solve Eq.~\eqref{Teqt0} in the main text, obtaining Eq.~\eqref{eq:Txyt0}. Let us introduce the notation
\be
\mathcal{T}\left(x,y,0,k\right)=\begin{pmatrix}\mathcal{T}_{11} & \mathcal{T}_{12}\\[1mm]
\mathcal{T}_{21} & \mathcal{T}_{21}
\end{pmatrix}
\ee
for the entries of the matrix, with the $x,y,k$ dependence suppressed for brevity.
We solve Eq.~\eqref{Teqt0} separately for the two columns of the matrix. For the left column, Eq.~\eqref{Teqt0} reads
\bea
\partial_{x}\mathcal{T}_{11}&=&-\frac{ik}{2}\mathcal{T}_{11}-iv\left(x,0\right)\sqrt{ik}\mathcal{T}_{21} \, ,\\
\label{T21eq}
\partial_{x}\mathcal{T}_{21}&=&-i\delta\left(x\right)\sqrt{ik}\mathcal{T}_{11}+\frac{ik}{2}\mathcal{T}_{21} \, .
\eea
It is convenient to solve these equations separately in the two regions $x<0$ and $x>0$. The general solution reads
\bea
\mathcal{T}_{11}&=&\begin{cases}
A_{-}e^{-ikx/2}-i\sqrt{ik}B_{-}e^{-ikx/2}e^{iky}I_{v}\left(x,y\right), & x<0,\\[1mm]
A_{+}e^{-ikx/2}-i\sqrt{ik}B_{+}e^{-ikx/2}e^{iky}I_{v}\left(x,y\right), & x>0,
\end{cases}\\[2mm]
\mathcal{T}_{21}&=&\begin{cases}
B_{-}e^{ikx/2}, & x<0,\\[1mm]
B_{+}e^{ikx/2}, & x>0.
\end{cases}
\eea
In order to find the four coefficients $A_{\pm}$ and $B_{\pm}$ (that depend on $y$ and $k$), we use the following four conditions:
(i) $\mathcal{T}_{11}$ must be continuous at $x=0$.
(ii) $\mathcal{T}_{21}$ has a jump discontinuity at $x=0$, whose magnitude is found by integrating Eq.~\eqref{T21eq} over an infinitesimally small region around $x=0$:
\be
\left.\mathcal{T}_{21}\right|_{x=0^{+}}-\left.\mathcal{T}_{21}\right|_{x=0^{-}}=-i\sqrt{ik}\left.\mathcal{T}_{11}\right|_{x=0} \, .
\ee
(iii) $\left.\mathcal{T}_{11}\right|_{x=y}=1$.
(iv) $\left.\mathcal{T}_{21}\right|_{x=y}=0$.
The latter two conditions come from the boundary condition $\mathcal{T}(x,x,t,k )=I$. These conditions yield four linear equations on the coefficients, whose solution then gives the left column of the matrix $\mathcal{T}\left(x,y,0,k\right)$ as given in Eq.~\eqref{eq:Txyt0}.
For the right column of the matrix, the solution is very similar (replacing $\mathcal{T}_{11}\to\mathcal{T}_{12}$ and $\mathcal{T}_{21}\to\mathcal{T}_{22}$), the difference being that conditions (iii) and (iv) give way to $\left.\mathcal{T}_{12}\right|_{x=y}=0$ and $\left.\mathcal{T}_{22}\right|_{x=y}=1$, respectively.
Solving the resulting equations for the coefficients then gives the right column of the matrix $\mathcal{T}\left(x,y,0,k\right)$ as given in Eq.~\eqref{eq:Txyt0}.

\end{appendices}

\section*{Acknowledgments}

The research of E.B. and B.M. is supported by the Israel Science Foundation (grants No. 1466/15 and 1499/20, respectively).


\begin{thebibliography} {99}
\bibitem{DNLSE}
A familiar form of the  DNLSE is 
$i \partial_t \phi+\partial_{x}^2 \phi- i \partial_x\left(|\phi|^2 \phi\right)=0\,,$
where $\phi(x,t)$ is a complex function \cite{KN}. The MFT equations~(\ref{d1}) and (\ref{d2}) coincide with the equations for the real and imaginary parts of $\phi(x,t)$, 
but in imaginary time and space.
\bibitem{smallboundarylayers} This adiabatic perturbation theory breaks down at times very close to $t=0$ and $t=1$, but these small regions (boundary layers in times)
do not contribute to the action in the leading order that we are interested in.
\end{thebibliography}
\end{document}